\documentclass[10pt, journal, compsoc]{IEEEtran}

\usepackage{mystyle}

\toggletrue{complete-version}
\togglefalse{highlight-revision}

\def\doctitle{On the Effectiveness of\\Transfer~Learning for Code~Search}

\def\docauthors{Pasquale Salza, Christoph Schwizer, Jian Gu, Harald C. Gall}

\def\dockeywords{%
Code search, transfer learning, source code modeling, multimodal embeddings, StackOverflow, deep learning.
}

\StrSubstitute{\doctitle}{\\}{ }[\cleandoctitle]
\StrSubstitute{\dockeywords}{.}{}[\cleandockeywords]
\hypersetup{
	pdftitle={\cleandoctitle},
	pdfauthor={\docauthors},
	pdfkeywords={\cleandockeywords}
}

\DeclareAcronym{api}{
	short = API,
	long = Application Program Interface
}

\DeclareAcronym{bert}{
	short = BERT,
	long = Bidirectional Encoder Representations from Transformers
}

\DeclareAcronym{ast}{
	short = AST,
	long = Abstract Syntax Tree
}

\DeclareAcronym{bpe}{
	short = BPE,
	long = Byte-Pair Encoding
}

\DeclareAcronym{cfg}{
	short = CFG,
	long = Control Flow Graph
}

\DeclareAcronym{dcg}{
	short = DCG,
	long = Discounted Cumulative Gain
}

\DeclareAcronym{ir}{
	short = IR,
	long = Information Retrieval
}

\DeclareAcronym{lstm}{
	short = LSTM,
	long = Long Short-Term Memory
}

\DeclareAcronym{mcm}{
	short = MCM,
	long = Masked Source Code Modeling
}

\DeclareAcronym{mlm}{
	short = MLM,
	long = Masked Language Modeling
}

\DeclareAcronym{mem}{
	short = MEM,
	long = Multimodal Embedding Model
}

\DeclareAcronym{mrr}{
	short = MRR,
	long = Mean Reciprocal Rank
}

\DeclareAcronym{ndcg}{
	short = NDCG,
	long = Normalized Discounted Cumulative Gain
}

\DeclareAcronym{nlp}{
	short = NLP,
	long = Natural Language Processing
}

\DeclareAcronym{nlpred}{
	short = NLPred,
	long = Next Line Prediction
}

\DeclareAcronym{nmt}{
	short = NMT,
	long = Neural Machine Translation
}

\DeclareAcronym{nsp}{
	short = NSP,
	long = Next Sentence Prediction
}

\DeclareAcronym{rnn}{
	short = RNN,
	long = Recurrent Neural Network
}

\DeclareAcronym{cnn}{
	short = CNN,
	long = Convolutional Neural Network
}

\DeclareAcronym{tf-idf}{
	short = tf-idf,
	long = term frequency–-inverse document frequency
}

\DeclareAcronym{anova}{
	short = ANOVA,
	long = ANalysis Of VAriance
}

\DeclareAcronym{qaa}{
	short = Q\&A,
	long = Question and Answer,
	plural-form = Questions and Answers
}

\DeclareAcronym{lamb}{
	short = LAMB,
	long = Layer-wise Adaptive Moments Based
}

\DeclareAcronym{cp}{
	short = CP,
	long = {Continuous Pretraining}
}

\DeclareAcronym{if}{
	short = IF,
	long = {Intermediate Finetuning}
}

\begin{document}

\title{\doctitle}

\author{
	Pasquale Salza, Christoph Schwizer, Jian Gu, and Harald C. Gall
	
	\IEEEcompsocitemizethanks{ %
		\IEEEcompsocthanksitem The authors are with the University of Zurich, Zurich, Switzerland. E-mail: \href{mailto:salza@ifi.uzh.ch}{salza@ifi.uzh.ch}, \href{mailto:christoph@schwizer.dev}{christoph@schwizer.dev}, \href{mailto:gu@ifi.uzh.ch}{gu@ifi.uzh.ch}, \href{mailto:gall@ifi.uzh.ch}{gall@ifi.uzh.ch}.
	}
}

\IEEEtitleabstractindextext{
	\begin{abstract}
The Transformer architecture and transfer learning have marked a quantum leap in natural language processing, improving the state of the art across a range of text-based tasks.
This paper examines how these advancements can be applied to and improve code search.
To this end, we pre-train a BERT-based model on combinations of natural language and source code data and fine-tune it on pairs of StackOverflow question titles and code answers.
Our results show that the pre-trained models consistently outperform the models that were not pre-trained.
In cases where the model was pre-trained on natural language \enquote{and} source code data, it also outperforms an information retrieval baseline based on Lucene.
Also, we demonstrated that the combined use of an information retrieval-based approach followed by a Transformer leads to the best results overall, especially when searching into a large search pool.
Transfer learning is particularly effective when much pre-training data is available and fine-tuning data is limited.
We demonstrate that natural language processing models based on the Transformer architecture can be directly applied to source code analysis tasks, such as code search.
With the development of Transformer models designed more specifically for dealing with source code data, we believe the results of source code analysis tasks can be further improved.
\end{abstract}


}
\maketitle

% Watermark
\watermark{%
This is the authors' version of the paper that has been accepted for publication in the\\%
IEEE Transactions on Software Engineering (TSE)%
}

\section{Introduction}
\label{sec:introduction}

\emph{Code search}, or \emph{code retrieval}, is the task of retrieving source code from a large code corpus given a natural language user query and can be an effective tool for software developers.
It helps them to find examples of how to implement a particular feature quickly, discover software libraries that provide specific functionality, navigate through their codebase, or even find pieces of source code that need to be changed to accommodate user concerns such as feature requests or bug fixes~\cite{ye_word_2016, palomba_recommending_2017}.
For example, a developer might search for \query{how to convert string to int in java} and the retrieval system returns a code snippet such as \code{int i = Integer.parseInt(intString);}.

The goal of code search is to return source code snippets that are most relevant to the user query.
In other words, the semantics of source code should correspond to the semantics of the natural language query.
Traditional retrieval systems are based on token matching, comparing the tokens in the search query with the tokens in the search corpus' documents and returning those documents with the biggest overlap between query and document tokens, which are often weighted by their frequency and inverse document frequency, or \emph{\acs{tf-idf}}~\cite{manning_introduction_2008}.
This approach has proven helpful for matching natural language queries with natural language documents, such as books or web pages.
However, when it is used to match natural language queries with source code documents is less effective.
One reason for this is that the tokens in the query do not necessarily match those in the source code.
For example, the query \query{read json data} would not find a method called \code{deserializeObjectFromString} even though it might be relevant to the query.
This discrepancy between the query language and the language in the documents of the search corpus is referred to as \enquote{lexical gap}, or \enquote{heterogeneity gap}.

Recent work has used neural networks~\cite{lv_codehow_2015, gu_deep_2018, shuai_improving_2020}, \ie, deep learning, to overcome the lexical gap, many of which have their origins in the field of \ac{nlp}.
Using models designed initially for \ac{nlp} tasks and applying them to problems dealing with source code can indeed be a viable approach since source code shows similar statistical properties as natural language~\cite{hindle_naturalness_2012}.
The model needs to understand the relationships between the tokens in the sequence for both natural and programming languages.
For natural language, that can mean finding the noun to which a pronoun refers or the subject to which a verb belongs.
In contrast, for source code, it might mean identifying opening and closing parentheses or matching variable access statements with inconsistent declaration statements.

The \transformer architecture has proven highly effective in modeling such dependencies between tokens, especially in longer sequences, where \acp{rnn} show some limitations~\cite{vaswani_attention_2017, devlin_bert_2019, radford_improving_2018}.
One aspect that makes them particularly powerful in \ac{nlp} tasks is the use of \emph{transfer learning}.
The idea behind it is to leverage a large corpus of data to \emph{pre-train} a model, and then \emph{fine-tune} it on a smaller dataset.
Commonly, the pre-training dataset is extensive, easy to acquire, but unlabeled, and not closely related to the problem we want to solve.
On the other hand, the fine-tuning dataset is generally characterized by being small, difficult to acquire, but often labeled and closely related to our problem task.
The intuition behind transfer learning is that, during pre-training, the model learns valuable abstractions of the data, which are effective for solving the problem, or \emph{\enquote{downstream}} task, during the fine-tuning.
In \ac{nlp}, pre-training usually consists of learning a language model on large corpora of natural language text.
Then, this pre-trained model can be employed in any particular downstream task, \eg, machine translation, sentiment analysis, part-of-speech tagging, and summarization.
An example of \transformer is \ac{bert}~\cite{devlin_bert_2019}, widely used for many \ac{nlp} tasks.
It has robustly optimized in many versions, \eg, \textsc{RoBERTa}~\cite{liu_roberta_2019}, and specialized~\cite{feng_codebert_2020}.
Moreover, \bert effectively enables transfer learning.

We argue that the same method can be applied to code search: train a language model on a large, unlabeled source code corpus, then fine-tune it on a smaller but labeled code search dataset.
The goal of this work is to leverage the predictive capabilities of \bert, as a state-of-the-art \transformer-based \ac{nlp} model and make use of transfer learning to improve the performance of code search.
We propose an approach based on pre-training two \bert encoders, one for queries and one for code, which learn how to independently represent those two forms of data.
Then, we assemble the encoders into a single \ac{mem}, fine-tuned on the code search downstream task.

Transfer learning can be beneficial in code search as attaining a large enough code search dataset for training is difficult.
Not only is a small dataset problematic for the training but also for the evaluation of a model since it limits the number of examples on which the model can be tested.
In our approach, we evaluate the performance of our models by leveraging both \github and \stackoverflow datasets that we specifically mined for this purpose.
We use the \github dataset to pre-train the \bert models, for a total of $\num{\approx 6450000}$, $\SI{\approx 27}{\times}$ larger than the \stackoverflow data.
We propose the use of \stackoverflow questions and accepted code answers as a proxy for code search interactions, whereby the question's title acts as an approximation of a search query and the code snippet of the accepted answer as the document to be retrieved from the search corpus.
We mine such a dataset for three popular programming languages, \ie, \javascript, \java, and \python, obtaining a total of $\num{\approx 240000}$ pairs of query and code.

To summarize, in this paper, we first define an approach for code search, using \transformers and transfer learning in the form of a \ac{mem}.
Therefore, we conduct a large empirical study and compare the produce model with the state-of-the-art approach for code search, \ie, \deepcs~\cite{gu_deep_2018} and an information retrieval-based approach, \ie, \lucene.
Finally, we also provide a combined approach based on the pre-filtering of search candidates by \lucene and then refined by \ac{mem}, and test its performance.

\begin{complete-version}
Our code search dataset from \stackoverflow, reflecting a typical transfer learning scenario, the pre-trained source code models, as well as the source code for data mining, pre-training, and fine-tuning, are available in our replication package~\cite{replicationpackage} and published at \url{https://tl-codesearch.netlify.app}.
\end{complete-version}

\paragraph{Paper organization}
The rest of the paper is structured as follows.
In \cref{sec:background}, we give an overview of the main concepts involved in this work.
\Cref{sec:approach} presents our approach based on \bert and transfer learning.
In \cref{sec:experimental_design}, we describe the experimental evaluation of our approach.
The results of the experiments are presented in \cref{sec:results}, whereas \cref{sec:related_work} surveys the related work. 
Finally, this paper concludes in \cref{sec:conclusions} with a summary of the findings and contributions of this work, as well as an outlook on future research in this area.

\section{Background}
\label{sec:background}

This section introduces the main involved concepts, which help understand the proposed approach.

\subsection{\Acf{mem}}
\label{subsec:background:mem}

An \acs{mem} builds vector representations (\enquote{embeddings}) for each mode, \eg, natural language and source code, such that similar concepts are located in the same region of a shared vector space, also called \enquote{semantic space}.
Recent work has relied on multimodal embeddings to overcome the lexical gap~\cite{guo_deep_2019,baltrusaitis_multimodal_2018}.
Multimodal embeddings are especially useful for code search as they allow for retrieval using a simple distance-based similarity metric, \eg, \enquote{cosine similarity}.
At search time, the natural language query is encoded into its vector representation and compared to all source code vectors in the search corpus.
Finally, the source code documents are returned as a list sorted by their distance to the query vector in increasing order.

\begin{figure}[tb]
	\centering
	\includegraphics[width=1.0\linewidth]{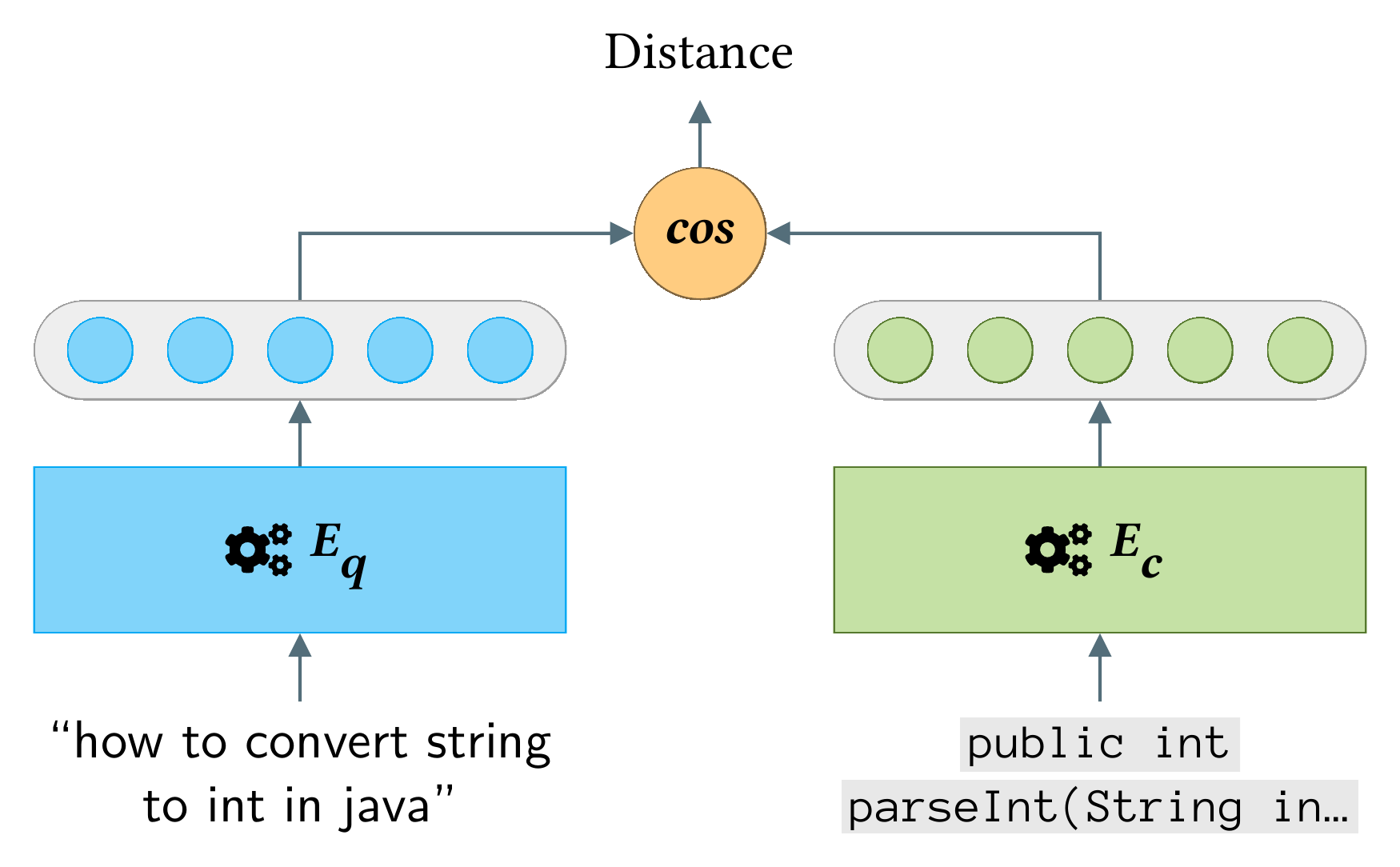}
	\caption{The \acf{mem}.}
	\label{fig:mem}
\end{figure}

To transform a natural language query into its vector representation, the \ac{mem} runs the query through an encoder $E_q$.
In contrast, another encoder $E_c$ transforms a source code document into its vector representation.
More formally, $E_q: \mathcal{Q} \rightarrow \mathbb{R}^d$ and $E_c: \mathcal{C} \rightarrow \mathbb{R}^d$ are embedding functions, where $\mathcal{Q}$ is the set of natural language queries, $\mathcal{C}$ is the set of source code documents, and $\mathbb{R}^d$ is the space of real-valued vectors of size $d$.
\Cref{fig:mem} depicts the architecture of a typical \ac{mem} for code search.
The encoder can be any model that converts the input data into its vector representation.
In the past, \acp{rnn} were often used for the source code encoder~\cite{guo_deep_2019, iyer_summarizing_2016, hussain_deep_2020, karampatsis_maybe_2019, hussain_codegru_2020, gu_deep_2018} as well as \acp{cnn}~\cite{shuai_improving_2020}.
In this work, we use \bert~\cite{devlin_bert_2019} as the encoder architecture for both source code and queries.

\subsection{\Acf{bert}}
\label{subsec:background:bert}

When the \transformer architecture was introduced, it replaced \acp{rnn} as the state of the art in \ac{nmt}~\cite{vaswani_attention_2017}.
\Acp{rnn} process each token in a sequence in a \enquote{sequential} way.
This leads to a loss of information on far-away tokens, \ie, by the time the \ac{rnn} arrives at the last token, the signal from the first token becomes very small.
The \enquote{attention} mechanisms mitigate this problem, allowing the \ac{rnn} to focus on arbitrary preceding tokens in the sequence.
Despite \emph{attention}, the nature in which \acp{rnn} process data is still sequential.
The \transformers change this by removing recurrence and handling the entire input sequence in parallel.
It achieves this by relying solely on attention, whereby a weight for each token pair in the input sequence is calculated.
This component is called an \enquote{attention head} and lets the model represent relationships between tokens in the sequence.
In fact, it has been shown that attention heads learn syntactic features of a (natural) language, such as prepositions and their corresponding object or nouns and their determiner~\cite{clark_what_2019}.

The parallel nature of \transformers facilitates faster training.
In turn, it enables training on much larger datasets, a key aspect that \bert~\cite{devlin_bert_2019} exploits, which was trained on an English corpus of \num{3.3} billion words.
The training tasks were \ac{mlm} and \ac{nsp}.
In \ac{mlm}, some tokens in the sequence are masked by a special \code{[MASK]} symbol, and the model has to predict the token that is masked out.
In \ac{nsp}, the model is given two random sentences from the corpus and has to decide if they appear in a sequence of one another.
These tasks guide \bert to learn a model of English.

\bert uses the same architecture as the \transformer with one distinction: while the \transformer employs an encoder-decoder architecture, \bert only uses encoders.
\bert does not generate an output sequence and is designed only to analyze the input sequence.
What sets \bert apart from similar \transformer-based models is its bidirectionality.
As opposed to Radford \etal's GPT model~\cite{radford_improving_2018} that processes the input sequence only in one direction, \eg, from left to right, \bert handles the input sequence in both directions (also from right to left) simultaneously.
It enhances the capabilities of the attention heads as they can focus on preceding and subsequent tokens.
Consequently, the token embeddings that \bert creates are dependent on the surrounding tokens and therefore called \enquote{\emph{contextualized} embeddings.}
They are more capable than context-free embeddings, \eg, those generated by word2vec~\cite{mikolov_distributed_2013} or GLoVe~\cite{pennington_glove_2014}, as they can distinguish between words that spell the same but have a different meaning, \eg, \query{minute} in \query{she pays attention to every minute detail} vs. \query{he was one minute late}.

\subsection{Transfer Learning}
\label{subsec:background:transfer_learning}

A language model can be practical in itself, \eg, it can be used to give typing suggestions~\cite{chen_gmail_2019}.
However, in the case of \bert, the language modeling tasks were only used as parameter initialization for different training tasks, such as question answering and language inference.
Transfer learning is a paradigm of \enquote{transferring} the knowledge learned from base data, usually a large dataset, to the new data for a new given domain or different tasks~\cite{pan_survey_2010, niu_decade_2020, georgekarimpanal_selforganizing_2018}.
The standard methodology consists of \enquote{pre-training} a model on a large corpus of unlabeled data and then \enquote{fine-tuning} it on a smaller supervised dataset.
During the pre-training stage, the model is usually trained in a \enquote{self-supervised} learning fashion, for which the unlabeled data is sufficient for the objective.
Thus, the pre-training data is usually extensive and readily available.
Instead, during the fine-tuning stage, the model is trained in a \enquote{supervised} learning way for which a ground truth is required.
The data used for the downstream task is supervised, and its quality matters the most.
Therefore, the data amount is usually limited since the cost to collect it is more expensive.
Other than classical transfer learning, there are other proposed strategies to improve data adaptability to target domains or tasks, which mainly work between the usual pre-training and fine-tuning phases.

\Ac{cp} is defined as tailoring a model to another data domain or designated task through a second phase of pre-training~\cite{gururangan_don_2020}, and it can lead to performance gain.
Moreover, multiphase adaptive pre-training, \eg, domain-adaptive training followed by the task-adaptive one, promises an even larger gain.
Instead, \ac{if} uses a pre-trained model and introduces intermediate tasks during an additional training stage, in the form of warm-up before training for the target task~\cite{pruksachatkun_intermediatetask_2020}.
Intermediate tasks can be of different levels of difficulty.
The simple intermediate tasks are close to learning the low-level skills, such as preserving the raw content and detecting the shallow attributes, \eg, verb tenses or sentences length in the case of \ac{nlp}.
In contrast, complex intermediate tasks are generally rather beneficial to promote the model, \eg, natural language inference~\cite{storks_recent_2019}, and question answering~\cite{min_knowledge_2019}.
Thus, they expect the model to have strong capabilities such as perceiving interrelations.

In this work, we use a \enquote{classical} transfer learning strategy to study its feasibility when applied to code search.
However, other strategies remain an important future work.

\section{Approach}
\label{sec:approach}

\begin{revised}
\begin{figure*}[tb]
	\centering
	\includegraphics[width=1.0\linewidth]{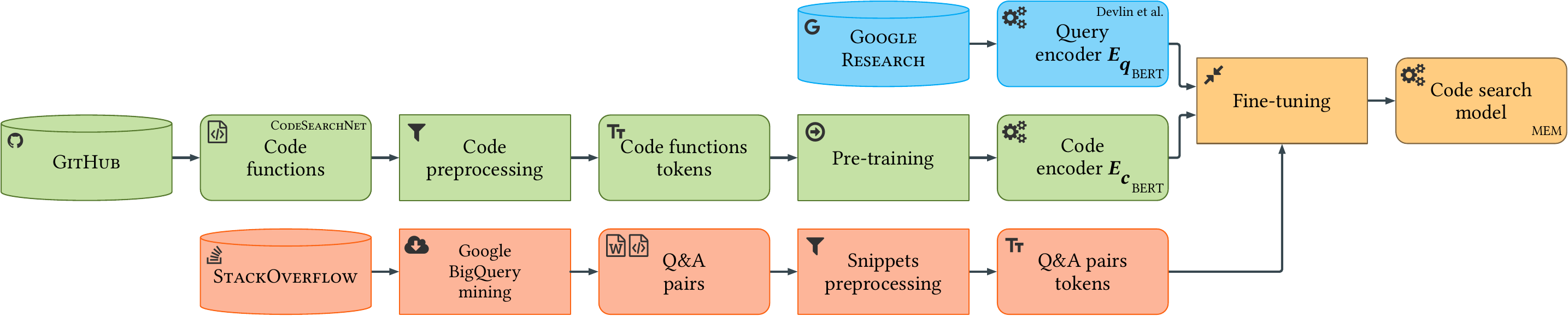}
	\caption{\begin{revised}The workflow of the approach composition, from data mining to final model \acs{mem} building.\end{revised}}
	\label{fig:workflow}
\end{figure*}

The approach proposed in this paper consists of two pre-trained \bert-based encoders, one for queries and one for code, ultimately assembled to an \ac{mem} model fine-tuned on \stackoverflow \acp{qaa} pairs data.
\Cref{fig:workflow} depicts the workflow describing the whole approach composition.
In this section, we describe such a proposed approach, giving details on the query and code encoders pre-training, \stackoverflow \acp{qaa} pairs data mining, and final \ac{mem} fine-tuning.
\end{revised}

\subsection{Pre-Training}
\label{subsec:approach:pre_training}

The first two pipelines in \cref{fig:workflow} represent the creation of the encoders used in our approach, namely the query $E_q$ and code $E_c$ encoders, using the so-called \enquote{pre-training} method.
During the pre-training, the models learn to represent their input into vectors, queries, and code for $E_q$ and $E_c$, respectively.
The two encoders will be lately assembled and re-trained, \ie, fine-tuning, for the task of code search.

In the case of the query encoder, \ie, $E_q$, we use the Devlin \etal's~\cite{devlin_bert_2019} pre-trained English model (uncased).
The model is publicly available~\cite{bertbase} and was trained by using \textsc{BookCorpus}~\cite{zhu_aligning_2015}, a dataset of \num{11038} and \textsc{Wikipedia}~\cite{englishwikipedia} (documents in English, excluding lists, tables, and headers).
Since the query encoder is already pre-trained, and we did not further modify it, we focus on the description of the code encoder $E_c$ pre-training.

\paragraph{Data collection}
We chose the \codesearchnet~\cite{husain_codesearchnet_2019} dataset, which was mined from \github repositories and consists of function definitions across six different programming languages (\javascript, \java, \python, \php, \go, and \ruby).
We decided on the \codesearchnet dataset because it readily provides a large set of source code samples in a machine-readable format.
To reduce natural language occurrences in the data, all documentation and comments were removed using a parser, namely \textsc{Tree-sitter}~\cite{treesitter}.
Otherwise, the data was not further processed.

\begin{table}[tb]
	\caption{Size of the pre-training datasets}
	\label{tab:pretrain_distribution}
	\centering
	% \resizebox{0.85\linewidth}{!}{
	\rowcolors{2}{}{gray!10}
\begin{tabular}{
    l S[table-format=7] S[table-format=9]
}

\hiderowcolors
\toprule
\textbf{Language} & {\textbf{Number of Functions}} & {\textbf{Number of Tokens}} \\

\midrule
\showrowcolors

\tabbvalue \javascript & \tabbvalue 1857835 & \tabbvalue 128430003 \\
\tabbvalue \java & \tabbvalue 1569889 & \tabbvalue 75654447 \\
\tabbvalue \python & \tabbvalue 1156085 & \tabbvalue 50551794 \\
\php & 977821 & 53352522 \\
\go & 726768 & 37075579 \\
\ruby & 164048 & 5495442 \\

\midrule

\tabbvalue \toplangs & \tabbvalue 4583809 & \tabbvalue 254636244 \\
\tabbvalue \all & \tabbvalue 6452446 & \tabbvalue 350559787 \\

\bottomrule

\end{tabular}

	% }
\end{table}

\Cref{tab:pretrain_distribution} lists the size of our pre-training dataset.
In addition to the dataset sizes of the individual languages, the table lists the combined size of all datasets (\all), as well as the combined size of the three largest datasets (\javascript, \java, and \python) (\toplangs).
We pre-trained the models on the \javascript, \java, and \python datasets, the \toplangs dataset, and the \all dataset.
To keep the number of experiments attainable, we forwent pre-training on the smaller \php, \go, and \ruby datasets.
Our largest dataset (\all) contains around \num{350} million tokens.
In comparison, \bert~\cite{devlin_bert_2019} was pre-trained on a corpus of \num{3.3} billion words (\num{0.8} billion words from the \textsc{BookCorpus}~\cite{zhu_aligning_2015} and \num{2.5} billion words from English \textsc{Wikipedia}~\cite{englishwikipedia}).

\paragraph{Configuration}
The pre-training procedure on source code is similar to the one by Kanade \etal~\cite{kanade_learning_2020} and identical to the pre-training of Devlin \etal's \bertbase model on natural language~\cite{devlin_bert_2019}, with only a slight difference in the pre-training tasks.
Instead of the \acf{nsp} task for pre-training on natural language, for source code, we apply \ac{nlpred}.
In this binary classification task, the model has to decide for any two given lines of source code \code{A} and \code{B}, whether \code{B} appears directly after \code{A}.
To train the model on this task, it is fed with samples from our pre-training dataset, in which \SI{50}{\percent} of the line \code{B} follows line \code{A}.
In the other \SI{50}{\percent} of the cases, \code{B} is a randomly chosen line from the corpus and does not immediately follow line \code{A}.
It is worth noting that \emph{no} extensive normalization practices were applied to the code before processing it.
In particular, since \ac{nlpred} is based on the concept of new lines, source code might present a splitting not correspondent to a logical splitting, \ie, single statements that are split over multiple lines.
A formatter might potentially be able to normalize the code beforehand.
However, we decided to keep the code as it is, except for removing empty lines.
First, the use of formatters might require the code to be fully parsable, which is not entirely guaranteed.
Instead, if based on regular expressions, they might also introduce errors.
Second, our downstream task is optimizing code search using \stackoverflow snippets data.
In the majority of the cases, code snippets are not valid code~\cite{terragni_csnippex_2016, terragni_apization_2021}, therefore their normalization would not be guaranteed by the use of formatters.
Therefore, keeping the code intact also makes the final model learn to deal with actual splitting styles.

We call the \ac{mlm} task for source code as \acf{mcm} to remark that the model is pre-trained on source code data instead of natural language.
Other than that, the \ac{mlm} and \ac{mcm} tasks are identical, \ie, the model has to predict masked out tokens in the input sequence.
Like Devlin \etal, we selected \SI{15}{\percent} of the tokens in the input sequence for masking.
In contrast, we only used a maximum sequence length of \num{256} tokens, whereas Devlin \etal used \num{512}.
The reason for this is that longer sequences require exponentially more memory during training and would thus not have fit in our GPU memory (see \cref{subsec:experimental_design:execution_setup}) without a drastic reduction in batch size.
Moreover, as we will see in \cref{tab:so_quality}, the average sequence length in our fine-tuning dataset is less than \num{256}, so most of the samples can be encoded by our model in their entirety.
It is worth noting that longer sequences are instead truncated.
With a sequence length of \num{256}, the maximum batch size fitting in our GPU memory was \num{62}.

Like Devlin \etal, we tokenized the source code sequence using \textsc{WordPiece} tokenization~\cite{wu_google_2016} with a vocabulary size of \num{30522} tokens.
\textsc{WordPiece}~\cite{schuster_japanese_2012} is one of the most used subword-based tokenization algorithms, which increased its popularity thanks to \ac{bert}.
The algorithm initializes the vocabulary with all the characters in the language, then iteratively combines pieces by maximizing the likelihood of the training data once added to the vocabulary.
Moreover, similar to Husain \etal~\cite{husain_codesearchnet_2019}, we kept the case information.
We second their choice to treat source code case-sensitively as case information carries a valuable signal, such as distinguishing between constants and variables or between class and method declarations.
In order to adapt \ac{bert} to the source code context, we ran \textsc{WordPiece} on top of the training data.

Devlin \etal pre-trained their model for \num{1} million steps, which equals about \num{40} epochs on their dataset.
Since our pre-training datasets are much smaller, and we used a different batch size and different sequence lengths, we adjusted the number of training steps accordingly to train for about \num{40} epochs as well.
For example, our \javascript dataset consists of \num{128430003} tokens.
With a sequence length of \num{256} tokens and a batch size of \num{62} sequences there are \num{15872} tokens in a batch.
Thus, we reach \num{40} epochs after pre-training for \num{323665} steps ($\num{323665} \times \num{15872} / \num{128430003}$).
Because of the smaller number of training steps, we also reduced the number of warm-up steps.
\Cref{tab:pretrain_params} lists the hyperparameters we used for pre-training.
We achieved high accuracy values for both tasks, \ie, above \SI{86}{\percent} on \ac{mcm} and \SI{95}{\percent} on \ac{nlpred}, which suggest that pre-training was successful, and the models learned useful abstractions of source code.

\begin{table}[tb]
	\caption{Pre-training hyperparameters vs. \bertbase\cite{devlin_bert_2019}}
	\label{tab:pretrain_params}
	\centering
	% \resizebox{0.8\linewidth}{!}{
	\rowcolors{2}{}{gray!10}
\begin{tabular}{
    l S[table-format=5.4] S[table-format=5.4]
}

\hiderowcolors
\toprule
\textbf{Parameter} & {\textbf{\bertbase}} & {\textbf{\bertcustom}} \\

\midrule
\showrowcolors

Optimizer & {Adam} & {Adam} \\
Learning rate & 0.0001 & 0.0001 \\
$\beta_1$ & 0.9 & 0.9 \\
$\beta_2$ & 0.999 & 0.999 \\
L2 weight decay & 0.01 & 0.01 \\
Learning rate decay & {linear} & {linear} \\
Dropout probability & 0.1 & 0.1 \\
Activation function & {gelu} & {gelu} \\
Masking rate & 0.15 & 0.15 \\
Hidden size & 768 & 768 \\
Intermediate size & 3072 & 3072 \\
Attention heads & 12 & 12 \\
Hidden layers & 12 & 12 \\
Vocabulary size & 30522 & 30522 \\
Maximum sequence length & 512 & \tabhvalue 256 \\
Batch size & 256 & \tabhvalue 62 \\
Learning rate warmup steps & 10000 & \tabhvalue 1000 \\

\bottomrule

\end{tabular}

	% }
\end{table}

\subsection{Query and Code Pairs Mining}
\label{subsec:approach:mining}

While, with the \codesearchnet dataset~\cite{husain_codesearchnet_2019}, we had a large enough dataset for pre-training, we needed a different dataset for fine-tuning.
We could have used the same dataset for both pre-training and fine-tuning, but not only would that have reduced the amount of data available for each phase, but also it would not reflect a typical transfer learning scenario in which the pre-training dataset \emph{differs} from the fine-tuning dataset.
We believe that method-docstring data, of which the \codesearchnet dataset consists, is not well suited for simulating code search because docstrings are very different from code search queries.
Not only are they usually much longer than search queries, but they are also commonly formulated only \emph{after} the code has been written.
The latter is fundamentally different from a search query formulation, where, typically, the query is formulated without prior knowledge of what a relevant search result appears.

Therefore, we decided to mine our dataset of question-answer pairs from \stackoverflow.
We use the question's title as the natural language query and the answer's code snippets as the source code document to be retrieved from the search corpus.
We believe that \stackoverflow questions are a good proxy for search queries, primarily since the platform is mostly used for finding code solutions.
Additionally, using \stackoverflow data allows us to build a large enough dataset to fine-tune and evaluate our models, which would have been very difficult to achieve with human annotations only.
We deliberately use only the question's \emph{title} and ignore the question's more detailed description.
We can thus ensure that the pre-training data is different from the fine-tuning data (following a typical transfer learning scenario) and that the natural language examples, \ie, question titles, resemble search queries sent to a code search engine.

\paragraph{Data extraction}
We extracted the data from \stackoverflow by using \textsc{Google BigQuery}~\cite{google_bigquery}, which contains an updated version of the \stackoverflow data dump, with the convenient availability of SQL functionalities.
\begin{complete-version}
The detailed SQL queries are available in our replication package~\cite{replicationpackage}.
\end{complete-version}
To gather examples that are specific to a programming language, we filtered questions by \query{javascript}, \query{java}, and \query{python} tags.
To gather more data, we included partial matches as well, which resulted in questions with tags such as \query{javascript-framework}, \query{javabeans}, or \query{python-3.6} to be part of our corpus.

\paragraph{Data quality improvement}
We selected only question-answer pairs whose answer was an accepted answer.
Since only the question poster can mark an answer as \enquote{accepted,} we can assume that an accepted answer reflects the solution for which the question poster was looking.
Practically, the question poster finds that answer relevant to their question, which is the behavior we expect of a search engine: returning relevant results to the user's query.
We could have selected the highest upvoted answer to build question-answer pairs.
However, since every \stackoverflow user can upvote an answer, we do not know anything about the relevance of that answer regarding the poster's intent.
We believe that accepted answers are better than the highest upvoted answers for a code search application to build question-answer pairs.

Like Husain \etal~\cite{husain_codesearchnet_2019}, we filtered out code answers that have fewer than three lines of code as these are pretty noisy.
Many of them contain only library import statements, or they have code that is not written in the target programming language, such as SQL queries, regular expressions, or command-line instructions.
Moreover, to further increase the quality of our sample, we removed any question-answer pairs in which either the question or the answer received fewer than three upvotes.
It allowed leveraging the crowd information that \stackoverflow offers since other users manually judged the relevance of a certain answer to the poster's intent.

Overall, our data mining process included the following steps:
\begin{inparaenum}[(1)]
    \item filter \stackoverflow questions by \query{javascript}, \query{java}, and \query{python} tags;
    \item remove questions that do not have an accepted answer;
	\item remove questions whose accepted answer does not contain a code snippet (using the \code{<pre><code>} tags);
	\item concatenate several code snippets of the same answer into one;
	\item discard text outside \code{<pre><code>} tags;
    \item remove question-answer pairs where either the question or the answer has fewer than three upvotes or where the answer contains fewer than three lines of code.
\end{inparaenum}
Filtering all \stackoverflow questions by the \query{javascript}, \query{java}, and \query{python} tags resulted in about \num{2} million \javascript-, \num{1.8} million \java-, and \num{1.8} million \python-related questions, of which roughly half had an accepted answer.
After having applied all the other steps, we were left with \num{85049} \javascript, \num{71194} \java, and \num{87231} \python question-answer pairs.
\Cref{tab:so_statistics} lists the number of samples remaining after each filtering step.

\begin{table}[tb]
	\caption{Number of StackOverflow questions after each filtering step. The numbers in the last row represent our final dataset sizes}
	\label{tab:so_statistics}
	\centering
	\resizebox{1.0\linewidth}{!}{
	\rowcolors{2}{}{gray!10}
\begin{tabular}{
    l S[table-format=7] S[table-format=7] S[table-format=7]
}

\hiderowcolors
\toprule
\textbf{Step} & {\textbf{\javascript}} & {\textbf{\java}} & {\textbf{\python}} \\

\midrule
\showrowcolors

Questions & 2045114 & 1841296 & 1884571 \\
Questions with accepted answer & 1105690 & 934062 & 984989 \\
Accepted answer contains a code snippet & 861273 & 533217 & 655430 \\

\midrule

$3+$ upvotes and $3+$ lines of code & \tabbvalue 85049 & \tabbvalue 71194 & \tabbvalue 87231 \\

\bottomrule

\end{tabular}

	}
\end{table}

When analyzing the effects of the last filtering step (see \cref{tab:so_quality}), we realize that, even though we only removed questions and answers with fewer than three upvotes, the average number of upvotes increased for each programming language by at least a factor of five for the questions, and at least a factor of four for the accepted answers.
Furthermore, while the average question length became slightly smaller, the average answer length became noticeably larger, both in the number of tokens and lines.

\begin{table}[tb]
	\caption{Dataset quality statistics on average after filtering}
	\label{tab:so_quality}
	\centering
	\resizebox{1.0\linewidth}{!}{
	\sisetup{table-format=3.2}
\rowcolors{2}{}{gray!10}
\begin{tabular}{
    l SS SS SS
}

\hiderowcolors
\toprule

\multirow{2}[2]{*}{\textbf{Statistic}} & \multicolumn{2}{c}{\textbf{\javascript}} & \multicolumn{2}{c}{\textbf{\java}} & \multicolumn{2}{c}{\textbf{\python}} \\
\cmidrule(lr){2-3} \cmidrule(lr){4-5} \cmidrule(lr){6-7}
& {\textbf{Before}} & {\textbf{After}} & {\textbf{Before}} & {\textbf{After}} & {\textbf{Before}} & {\textbf{After}} \\

\midrule
\showrowcolors

Question upvotes & 2.94 & \tabhvalue 21.16 & 3.18 & \tabhvalue 16.64 & 3.51 & \tabhvalue 18.37 \\
Question length (tokens) & 8.74 & 8.48 & 8.62 & 8.49 & 9.08 & 8.66 \\
Answer upvotes & 4.75 & \tabhvalue 28.96 & 5.14 & \tabhvalue 22.61 & 5.34 & \tabhvalue 24.13 \\
Answer length (tokens) & 175.61 & 207.39 & 203.50 & 262.99 & 165.15 & 205.90 \\
Answer length (lines) & 29.73 & 34.43 & 29.64 & 37.73 & 25.89 & 32.63 \\

\bottomrule

\end{tabular}

	}
\end{table}

\paragraph{Snippets cleaning}
We concatenated all code snippets for answers containing more than one code snippet into one (separated by a newline character).
For instance, an answer author might have alternated their response with code snippets and explanatory text.
Then, we removed all text not contained in the code snippets.

Sometimes, the snippet answer contains code comments that further explain it.
It should be noted that comments were explicitly removed from the pre-training data using a parser.
Since the code snippets in \stackoverflow answers are not necessarily syntactically correct, we cannot use a parser to remove comments from the answer snippets.
Unfortunately, it is known that the code snippets are, in general, hardly statically parsable since diverging from a well-formed shape~\cite{terragni_csnippex_2016, subramanian_making_2013}.
We could exclude non-parsable answers from the dataset, but not only would that reduce the size of our dataset, but also it is not necessary for our model to receive syntactically correct code since it is purely token-based.
It is one advantage over models that make use of syntactic structure in the code, such as \acp{ast}.
Nevertheless, the fact that our pre-trained model has not seen comments will likely affect its performance during fine-tuning.

\subsection{Fine-Tuning}
\label{subsec:approach:fine_tuning}

The fine-tuning procedure for code search closely follows the design by Husain \etal~\cite{husain_codesearchnet_2019}.
We use the same \ac{mem} architecture with two encoder models, one for the natural language queries and one for the source code snippets (see \cref{subsec:background:mem}), and the same training objective, namely reducing the distance, \ie, cosine distance, between query and code vector in the vector space.

\paragraph{Configuration}
For the fine-tuning of our \acl{mem}, we used the hyperparameters listed in \cref{tab:finetune_params}.
Since our fine-tuning procedure is very similar to the one by Husain \etal, we kept their hyperparameters whenever possible.
We increased the maximum sequence length of the code encoder to \num{256} because the average code snippet in our fine-tuning dataset has around \num{180} tokens (see \cref{tab:so_quality}) and because we pre-trained our code encoder with the same maximum sequence length of \num{256}.
We kept the maximum sequence length for the query encoder at \num{30} tokens as our average query contains only around \num{9} tokens.
Thus, we do not expect better performance with a larger sequence length.
To support a high number of experiments combinations and repetitions, we conducted some preliminary runs and observed a quicker convergence with \ac{lamb}~\cite{you_large_2019} than using Adam~\cite{kingma_adam_2015}.
We then used \ac{lamb} and limited training to \num{5} epochs.
In contrast, Husain \etal trained with Adam for a maximum of \num{500} epochs but applied early stopping, \ie, their training stopped if the \ac{mrr} did not improve for \num{5} epochs (\enquote{patience} hyperparameter).
\num{32} was the largest batch size fitting the memory of our Nvidia Tesla V100 (\SI{32}{\giga\byte}, see \cref{subsec:experimental_design:execution_setup}).

The \bert-specific hyperparameter values were mostly dictated by our pre-trained models.
For example, the English model provided by Devlin \etal was pre-trained on a vocabulary of \num{30522} tokens.
To keep the hyperparameters between the code and query encoder as similar as possible, we also pre-trained our source code model on a vocabulary size of \num{30522} tokens.
The same holds for the hidden size and the intermediate size.
The only hyperparameters we changed from our pre-trained models were the number of attention heads and the number of hidden layers (both had a value of \num{12} during pre-training).
We decided to use Husain \etal's values (\num{8} and \num{3}, respectively) because we observed faster convergence of the models during training with those values, presumably due to the reduced model complexity.

\begin{table}[tb]
	\caption{Fine-tuning hyperparameters vs. Husain \etal~\cite{husain_codesearchnet_2019}}
	\label{tab:finetune_params}
	\centering
	\resizebox{1.0\linewidth}{!}{
	\rowcolors{2}{gray!10}{}

\begin{tabular}{
    l S[table-format=5.2] S[table-format=5.2]
}

\hiderowcolors
\toprule
\textbf{Parameter} & {\textbf{Husain \etal}} & {\textbf{Our Approach}} \\

\midrule

\multicolumn{3}{l}{\textit{\acl{mem} hyperparameters}} \\

\midrule

\showrowcolors

Learning rate & 0.0005 & 0.0005 \\
Learning rate decay & 0.98 & 0.98 \\
Momentum & 0.85 & 0.85 \\
Dropout probability & 0.1 & 0.1 \\
Maximum sequence length (query) & 30 & 30 \\
Maximum sequence length (code) & 200 & \tabhvalue 256 \\
Optimizer & {Adam} & \tabhvalue {LAMB} \\
Maximum training epochs & 500 & \tabhvalue 5 \\
% Patience & 5 & \tabhvalue 5 \\
Batch size & 450 & \tabhvalue 32 \\

\midrule
\hiderowcolors

\multicolumn{3}{l}{\textit{\bert-specific hyperparameters (both code and query)}} \\

\midrule
\showrowcolors

Activation function & {gelu} & {gelu} \\
Attention heads & 8 & 8 \\
Hidden layers & 3 & 3 \\
Hidden size & 128 & \tabhvalue 768 \\
Intermediate size & 512 & \tabhvalue 3072 \\
Vocabulary size & 10000 & \tabhvalue 30522 \\

\bottomrule

\end{tabular}

	}
\end{table}

One difference between Husain \etal and our approach is the tokenization and vocabulary building process.
Because we used pre-trained models in our experiments, we had to use the vocabulary learned by the pre-trained models since the models' pre-trained weights depend on their specific encoding of tokens.
Husain \etal, on the other hand, did not rely on parameter weights of pre-trained models, which is why they built a new vocabulary from the fine-tuning data (the training set).
They used \ac{bpe}~\cite{sennrich_neural_2016} for that process, while the pre-trained English model (\bertbase) built its vocabulary using \textsc{WordPiece} tokenization~\cite{wu_google_2016}.
Both \ac{bpe} and \textsc{WordPiece} use subword information and work very similarly in creating the token vocabulary.
Hence, we do not expect the choice between \ac{bpe} and \textsc{WordPiece} tokenization to affect our results significantly.
Still, to keep things consistent in our experiments, we also used \textsc{WordPiece} tokenization to build our vocabulary.
For the pre-trained code models, the vocabulary was built from the pre-training data, while the non-pre-trained baseline models came from the training set of our fine-tuning data.

Like Husain \etal, we converted all query input to lowercase and kept the case information of the source code input.
The same is true for the pre-trained models.
We used the uncased version of Devlin \etal's English model~\cite{devlin_bert_2019} and pre-trained our source code models case-sensitively.

\section{Experimental Design}
\label{sec:experimental_design}

To examine the effectiveness of transfer learning for code search, we devised several experiments with different configurations for pre-training and fine-tuning of \acfp{mem}.
We use two distinct datasets to simulate a typical transfer learning scenario in which the pre-training data differs from the fine-tuning one.
The pre-training dataset consists of function definitions from open-source projects on \github, while the fine-tuning one contains \stackoverflow questions and corresponding code snippet answers.
In the context of the study, we formulate the following research questions.
\smallskip

\begin{reqs}
    \item [\req{1}] Does a pre-trained English natural language model improve the code search performance?
\end{reqs}
For a code search model to perform well, it needs to have a good understanding of the user's information need, which is expressed in the form of a natural language query.
To this end, we use a pre-trained English language model to examine whether the pre-training allows the code search model to learn better query representations and leads to better code search results.
\smallskip

\begin{reqs}
    \item [\req{2}] Does a pre-trained single-language source code model improve the code search performance?
\end{reqs}
For the code search model to retrieve a relevant code snippet from the search corpus for a given query, it has to build good representations of the source code snippets in the search corpus.
To answer this question, we pre-train a source code model on a specific programming language (\eg, \java), fine-tune and evaluate it on data of the same programming language (\java).
\smallskip

\begin{reqs}
    \item [\req{3}] Does a pre-trained English natural language model in combination with a pre-trained single-language source code model improve the code search performance?
\end{reqs}
This research question is the combination of research questions \req{1} and \req{2}.
The hypothesis is that if a pre-trained natural language model and a pre-trained source code model both lead to better code search performance, the combination of the two might lead to even better performance.
\smallskip

\begin{reqs}
    \item [\req{4}] Does a pre-trained multi-language source code model improve the code search performance?
\end{reqs}
For this purpose, we pre-train a source code model on several programming languages, fine-tune it, and evaluate it on a single programming language and a multi-language search corpus.
\smallskip

\begin{reqs}
    \item [\req{5}] Does the combination of an information retrieval method and transfer learning model improve the code search performance?
\end{reqs}
For this research question, we investigate the possibility of combining an information retrieval method, \ie, \lucene, with all the \acp{mem} investigated for the above research questions.
\smallskip

In the following, we describe the methodology we applied to answer the research questions mentioned above.
\begin{complete-version}
We provide all the details about the study in our replication package~\cite{replicationpackage}.
\end{complete-version}

\subsection{Methodology}

\begin{table}[tb]
	\caption{Approximate folds size after the \num{10}-fold split}
	\label{tab:evaluation_split}
	\centering
	% \resizebox{0.9\linewidth}{!}{
	\rowcolors{2}{}{gray!10}
\begin{tabular}{
    l
    S[table-format=6]
    S[table-format=6]
    S[table-format=5]
    S[table-format=5]
}

\hiderowcolors
\toprule
\textbf{Language} & {\textbf{Total}} & {\textbf{Training}} & {\textbf{Validation}} & {\textbf{Test}} \\

\midrule
\showrowcolors

\javascript & 85049 & 68889 & 7654 & 8504 \\
\java & 71194 & 57667 & 6407 & 7119 \\
\python & 87231 & 70657 & 7850 & 8723 \\
\toplangs & 243474 & 197213 & 21912 & 24347 \\

\bottomrule

\end{tabular}

	% }
\end{table}

To evaluate the models deriving from our approach, we apply \num{10}-fold cross-validation to all the experiments by splitting the entire dataset into ten equal folds and using nine for training and one for testing.
We further split the data from the nine training folds into \SI{90}{\percent} training and \SI{10}{\percent} validation data, leaving us with the fold sizes of \cref{tab:evaluation_split}.
With the number of observations from the cross-validation, we can apply statistical tests to mitigate the risk of spurious differences.
Since some question posts on \stackoverflow might be related to multiple programming languages, to avoid duplicates and ambiguities, we removed such intersections in the case of the \toplangs dataset.
This cleaning operation resulted into a removal of \num{1145} ($\SI{\approx 0.47}{\percent}$) pairs from \toplangs.

To test our models' performance, we apply two different strategies when evaluating:
\begin{itemize}[font=\normalfont\bfseries]
	\item[\evalonek] For each query in our test set, we search for the correct answer among \num{1000} code snippets (the correct code snippet and \num{999} distractor snippets), the evaluation strategy as adopted by Husain \etal~\cite{husain_codesearchnet_2019}.
	The distractor snippets are selected randomly from our test set.
	While a search corpus of \num{1000} code snippets is small, a fixed search corpus size makes our results uniformly comparable between different programming languages.
	\item[\evalfull] We use the full test set as a code snippets corpus for each of the queries to simulate a more realistic scenario in which developers could use such an approach.
\end{itemize}

As for evaluation measures, we use the \acf{mrr}, \topk{k}, and \aroma accuracy metrics, which are described in the following.

\subsection{Evaluation Metrics}
\label{subsec:experimental_design:evaluation_metrics}

Typical evaluation metrics for \ac{ir} are \emph{precision}, \emph{recall}, \emph{F-measure}, and \ac{dcg}.
These metrics only make sense if \emph{several} documents in the search corpus are relevant.
If, instead, there is precisely \emph{one} relevant (and known) document in the corpus, the reciprocal rank and \topk{k} accuracy are more suitable metrics.
We use these metrics to evaluate the performance of our approach, with the addition of the \aroma score metric to determine how good the choices that the techniques predict as an alternative are.

\paragraph{\Acf{mrr}}
The reciprocal rank is the inverse rank of the relevant document~\cite{wu_optimizing_2011}.
For instance, if the relevant document is returned at position \num{4}, the reciprocal rank is $1 / 4 = 0.25$, if it is returned at position \num{1}, the reciprocal rank is $1 / 1 = 1$.
The intuition behind the reciprocal rank is that if the relevant document appears at position $k$, the user must go through $k$ documents to find the relevant one.
At this point, the precision is $1/k$, which is also the reciprocal rank.
Finally, the \ac{mrr} is the average of multiple reciprocal ranks, \ie, from various queries.

\paragraph{Top-k accuracy}
The \topk{k} accuracy metric expresses how often, overall the evaluation samples, a predicted position of the document is within the first $k$ relevant documents~\cite{boyd_accuracy_2012}.
Applied to our context, the simple intuition behind this metric is to express how many documents a user has to read before finding the correct one.
We compute and report \topk{1}, \topk{3}, \topk{5}, and \topk{10} accuracy values.

\paragraph{\aroma-based similarity score}
The metrics mentioned above help understand how well a model performs to rank the expected code snippet associated with a specific \stackoverflow query.
However, we cannot exclude that other code snippets might be legitimately associated with multiple \stackoverflow titles, even if they do not belong to the same post.
Therefore, a model might potentially rank as first a code snippet that is not correspondent to the ground truth but, at the same time, represents a good match for the given query.
In practice, we need a way to establish how good are the models in identifying alternative solutions to the oracle.

For this reason, we employ an evaluation metric based on \aroma~\cite{luan_aroma_2019}, a tool for code-to-code similarity tool considering the structural aspects of source code.
In particular, \aroma was proved effective in identifying similarities between partial code snippets, \eg, obtained from \stackoverflow.
Similar to other contributions~\cite{cambronero_when_2019, li_neural_2019}, we use \aroma to define a metric for the similarity between the answers in our evaluation set.
This metric is intended to mimic the manual assessment of the correctness of search results but in an automatic and reproducible way~\cite{li_neural_2019}, without relying on human judgment that, considering the size of our dataset, would be infeasible.

The original \aroma implementation uses \textsc{ANTLR 4} to parse the source code and extract the structural features.
The only support available at the time of our experimentation was \java.
For this reason, we implemented a structural feature extractor for \python and \javascript.
Then, \aroma computes the number of overlapping structural features between a pair of snippets.
Such a number is the one we used as the basis for the \aroma-based similarity score we used in our experimentation.
The \aroma tool also applies other steps for pruning and clustering, but they are intended to be used for other purposes, \ie, code recommendation~\cite{luan_aroma_2019}.

Given a pair of snippets, \aroma returns an integer number.
We normalize such a value between \num{0.0} and \num{1.0} by using the following procedure.
Given a text query and actual value, \ie, the code snippet that is expected to be ranked as first, we retrieve the number of overlapping \aroma features between the true code and all the possible code snippets a compared model could choose as the best association.
We then rank these values, \ie, we transform the scores into rank positions, and apply a simple \emph{min-max normalization}, therefore resulting in values between \num{0.0} and \num{1.0}.
As a result, we can compute the \aroma similarity score between the expected snippet and the one the model selected as first for each instance of our test set.
For the sake of clarity, we refer to such a similarity score as \enquote{\aroma{}.}

\subsection{Compared Models}

We adopt a specific terminology for our experiments to identify the type of models to which we refer.
We use the following pattern to express the models we evaluate: \expt{$E_q$}{$E_c$}{Training}{Test}.
The pairs of brackets represent the different components of the models.
In particular, the curly brackets describe the dataset we used for pre-training.
The first part is for the query encoder, \ie, $E_q$, with a possible value as \exptlabel{NO}, meaning we initialize by random values the weights for the encoder, or \exptlabel{EN}, where we use the uncased version of Devlin \etal's English model~\cite{devlin_bert_2019, bertbase}.
Instead, the code encoder can be either \exptlabel{NO}, or one of the languages used for pre-training, \ie, \javascript (\exptlabel{JS}), \java (\exptlabel{JA}), \python (\exptlabel{PY}), \toplangs (\exptlabel{TP}), or \all (\exptlabel{AL}).
The square brackets represent the training component, \ie, fine-tuning: \exptlabel{NO}, \exptlabel{JS}, \exptlabel{JA}, \exptlabel{PY}, \exptlabel{TP}, and \exptlabel{AL}.
Finally, the round brackets represent the target search language, \ie, test, we use for the evaluation, with values: \exptlabel{JS}, \exptlabel{JA}, \exptlabel{PY}, and \exptlabel{TP}.

It is worth noting that some produced combinations correspond to some of the baselines we discuss in the following subsection.
Here, we describe the models that represent the main contribution of this work.
\begin{complete-version}
The complete list of experiments is published online in our replication package~\cite{replicationpackage}.
\end{complete-version}

\paragraph{Pre-trained query models (\req{1})}
First, we use Devlin \etal's~\cite{devlin_bert_2019} pre-trained English model \bertbase (uncased), which is publicly available~\cite{bertbase}, and applied it to the query encoder $E_q$.
It means that the weights of the query encoder were initialized with the weights of the pre-trained English model.
In this scenario, the code encoder $E_c$ is not pre-trained, \ie, its weights are initialized with random values.
The models used for comparison when we address this research question are expressed in the form: \expt{EN}{NO}{LANG}{LANG}.

\paragraph{Pre-trained code models (\req{2})}
Then, we use our own pre-trained source code models (see \cref{subsec:approach:pre_training}) to initialize the weights of the code encoder $E_c$.
This time, the weights of the query encoder $E_q$ are initialized with random values.
We limited the experiments to cases in which the pre-training is performed with the same programming language as the fine-tuning.
We note that cross-language learning, such as using a pre-trained \python model to fine-tune on \java data, could make sense in a scenario where the target language is so rare that there is not enough data available to justify pre-training.
However, we expect a pre-trained \emph{multi-language} source code model, \ie, a model that was trained on a mix of programming languages, to yield better results.
We examine multi-language source code models in \req{4}.
The models are expressed in the form: \expt{NO}{LANG}{LANG}{LANG}.

\paragraph{Pre-trained query and code models (\req{3})}
As a next step, we combine the pre-trained query and code models to see how they complement each other.
Both the weights of the query encoder $E_q$ and code encoder $E_c$ are restored from the respective pre-trained model.
Here, the models are expressed in the form: \expt{EN}{LANG}{LANG}{LANG}.

\paragraph{Pre-trained multi-language code models (\req{4})}
Afterward, we examine the source code models pre-trained on several programming languages.
We pre-trained two such models: one on \javascript, \java, and \python data (\toplangs) and another one on \javascript, \java, \python, \php, \go, and \ruby data (\all).
Again, we distinguish between only pre-training the query encoder $E_q$, only pre-training the code encoder $E_c$, and pre-training both.
For these experiments, in addition to the single-language datasets, we fine-tune and evaluate the models on a multi-language dataset consisting of \javascript, \java, and \python samples (\toplangs).

We pre-train two such models: one on \javascript, \java, and \python data (\toplangs) and another one on \javascript, \java, \python, \php, \go, and \ruby data (\all).
Again, we distinguish between only pre-training the query encoder $E_q$, only pre-training the code encoder $E_c$, and pre-training both.
For these experiments, in addition to the single-language datasets, we fine-tune and evaluate the models on a multi-language dataset consisting of \javascript, \java, and \python samples (\toplangs).

\subsection{Baselines}
\label{subsec:experimental_design:baselines}

\paragraph{Random}
First, we build a simple baseline that we call \random, since it is based on the random initialization of the weights for both the query and code encoders.
We do not apply any fine-tuning, and we compute the cosine distance on the target search language with the \enquote{random} encoders as they are.
The baseline is expressed in the form: \expt{NO}{NO}{NO}{LANG}.

\paragraph{Zero-shot}
Second, we evaluate all models without fine-tuning them.
This is often referred to in the literature as \emph{zero-shot learning}~\cite{socher_zeroshot_2013, xian_zeroshot_2017}.
We include this configuration as a baseline to estimate how useful the source code model is in itself, \ie, without any knowledge of the downstream task.
The term used for such a baseline is: \expt{$E_q$}{$E_c$}{NO}{LANG}.

\paragraph{No pre-train}
Then, we train the \ac{mem} \emph{without} any pre-training.
We use the same hyperparameters as in \cref{tab:finetune_params} to make our baseline comparable to the experiments with pre-trained models.
This baseline allows us to measure the effect of transfer learning, \ie, how much better the pre-trained models perform compared to a model trained from scratch.
We refer to this baseline as: \expt{NO}{NO}{LANG}{LANG}.

\paragraph{Information Retrieval (\lucene)}
We build a \lucene (v8.6.1) baseline with default parameters, as suggested by Hussain \etal~\cite{husain_codesearchnet_2019} (they mention \textsc{Elasticsearch}, which is based on \lucene).
\lucene is a widely used open-source search engine and retrieves documents using an inverted index structure and \ac{tf-idf} weighting between query and document.
By default, \lucene converts all text to lowercase and splits tokens based on grammar.
The intention behind this baseline is to give an estimate of what is possible with a low-effort and low-cost, \enquote{out-of-the-box} solution and to assess the usefulness of the \ac{mem}.
Note that the \lucene model does not require any training: it simply indexes all code snippets from the test set and retrieves them during evaluation.
We refer to the \lucene models with the string \exptlu{LANG}, where we only indicate the target search language.

\paragraph{\deepcs}
As for a comparison with existing approaches for code search based on neural networks, we executed the experiments by using \deepcs by Gu \etal~\cite{gu_deep_2018}, which we consider as the state of the art (see \cref{sec:related_work}).
We trained the \deepcs by using our data and producing a model for each programming language.
We refer to \deepcs models as: \exptdc{LANG}{LANG}.
The two parts correspond to the language used for training and test, respectively.
To have a fair comparison, we adapt some of the default configuration parameters of \deepcs.
In particular,
\begin{inparaenum}[(1)]
	\item we use a maximum sequence length for the code of \num{256}, instead of \num{50},
	\item a vocabulary size of \num{30522}, instead of \num{10000},
	\item a batch size of \num{32} instead of \num{64}.
\end{inparaenum}

\subsection{Combined Models (\req{5})}
\label{subsec:experimental_design:combined_models}

Finally, we produce a combination of an information retrieval method, \ie, \lucene, with all the models produced for the research questions mentioned above.
It is worth noting that, for \req{5}, we only consider the \evalfull evaluation strategy.
In particular, we build the combined model as a pipeline.
First, for each query, \lucene is used on the entire test set to establish the rankings.
Expressly, we set up a limited number of results to \num{1000}, which is the same number of samples used for the \evalonek evaluation strategy.
Second, when evaluating the \acp{mem}, we limit the choices between the \num{1000} samples that \lucene chose.
We can then consider \lucene to act as a sort of filter, reducing the number of samples between the \ac{mem} has to choose.
We refer to these models as \exptlumem{$E_q$}{$E_c$}{Training}{Test}, similarly to what we do with the \acp{mem}.

\subsection{Execution Setup}
\label{subsec:experimental_design:execution_setup}

Pre-training and fine-tuning were executed on a machine with an Intel Xeon Gold CPU clocked at \SI{2.60}{\giga\hertz}, \SI{16}{\giga\byte} RAM, and a single Nvidia Tesla V100 GPU with \SI{32}{\giga\byte} of memory.
Pre-training took between \num{1.6} and \num{11} days, depending on the size of the pre-training dataset.
Fine-tuning on a single fold (\num{5} epochs) took between \num{35} minutes and \num{2} hours, depending on the size of the fine-tuning dataset.

\subsection{Threats to Validity}

\paragraph{Internal validity}
The most significant limitation to our experimental design comes from the nature and quality of our evaluation dataset.
While using \stackoverflow questions and code answers allows us to gather large amounts of evaluation data, we cannot be sure that they are a valid proxy for measuring code search performance.
We may measure something else instead, such as how well our model can find the correct answer among multiple possible answers to a \stackoverflow question.

Furthermore, not all questions ask for a code answer to a concrete implementation problem.
Some questions touch on more high-level, abstract topics, such as programming style or best practices.
The answer to these questions may still contain code examples for demonstration purposes.
We observe a significant semantic discrepancy between the query and the corresponding code snippet in these cases.
Related to this is that code snippets alone might not give a comprehensive answer to the question posed, and it only makes sense in the context of the surrounding natural language explanations of the answer post.
This is especially true because, for answers that contain several code snippets, we concatenate them into one, which makes the code snippets less cohesive.
To mitigate this issue, we introduced the \aroma score as a metric to measure the relevance of the code snippets that the approaches classify as the correct answer (see \cref{subsec:experimental_design:evaluation_metrics}).

Additionally, the code snippets can contain comments, which we did not remove during pre-processing.
While we would want the comments to be included in the search results returned to the user, they may be considered noise to our code encoder, which was pre-trained on source code where comments were removed.
The same is true for console outputs, which are not removed from the evaluation dataset.

\paragraph{External validity}
Our results are limited in the way that they can be generalized to other source code analysis tasks.
While problems such as code summarization and code generation are very similar to code search, we did not evaluate those problem tasks experimentally.
This limitation is especially true because both those problems require \emph{generative} models that produce an output sequence (a natural language sequence in code summarization and a source code sequence in code generation).
The models we developed are only capable of finding code snippets from a corpus of \emph{existing} snippets.

Moreover, through our study, we cannot provide any insights on the type of information transferred, \eg, syntax or semantics.
However, recent research by Iyer \etal~\cite{sinha_masked_2021} suggests that the success of \ac{mlm} pre-training, as in the case of \bert, is most likely due to it learning higher-order distributional statistics that make for a useful prior for subsequent fine-tuning and not to its ability to discover syntactic and semantic mechanisms.
Affirming the same in the case of code processing is indeed an exciting and more sophisticated future work, requiring extended and specific experimentation.

\section{Results}
\label{sec:results}

In this section, we present the results from the experiments described in \cref{sec:experimental_design}.
We introduce the results of all experiments involving a single-language pre-trained model.
Then, we present the results of the pre-trained multi-language models.
We proceed with the analysis of the \topk{k} accuracy values trend.
Finally, we conclude with the analysis of the combined models, \ie, \lucene and \acp{mem}.
To compare the observations, we applied the \enquote{Kruskal-Wallis H} test~\cite{montgomery_design_2017}, and \enquote{Vargha-Delaney $\hat{A}_{12}$} test~\cite{vargha_critique_2000}, for the effect size to characterize the magnitude of such differences.

\begin{table*}[tb]
	\caption{Median values over \num{10} folds for all the computed metrics. We highlight in bold the maximum scores per language test set}
	\label{tab:results}
	\centering
	\resizebox{0.9\linewidth}{!}{

\sisetup{table-format=1.4, round-mode=places, round-precision=4}
\rowcolors{2}{gray!10}{}
\begin{tabular}{
    llr SSSSSS SSSSSS
}

\hiderowcolors
\toprule

\multirow{2}[2]{*}{\textbf{Language}} & \multirow{2}[2]{*}{\textbf{RQ}} & \multirow{2}[2]{*}{\textbf{Type}} & \multicolumn{6}{c}{\textbf{1K}} & \multicolumn{6}{c}{\textbf{Full}} \\
\cmidrule(lr){4-9} \cmidrule(lr){10-15}
& & & {\textbf{\ac{mrr}}} & {\textbf{\aroma}} & {\textbf{\topk{1}}} & {\textbf{\topk{3}}} & {\textbf{\topk{5}}} & {\textbf{\topk{10}}} & {\textbf{\ac{mrr}}} & {\textbf{\aroma}} & {\textbf{\topk{1}}} & {\textbf{\topk{3}}} & {\textbf{\topk{5}}} & {\textbf{\topk{10}}} \\

\midrule
\showrowcolors

\javascript & \req{1} & \exptlu{JS} & 0.237411 & 0.532144 & 0.1705 & 0.260188 & 0.306063 & 0.3695 & 0.132822 & 0.435978 & 0.0898883 & 0.14363 & 0.17231 & 0.214815 \\
& & \exptdc{JS}{JS} & 0.15663 & 0.404355 & 0.07525 & 0.16725 & 0.223875 & 0.325313 & 0.0446346 & 0.311378 & 0.0155791 & 0.0398589 & 0.0572016 & 0.0934156 \\
& & \expt{NO}{NO}{JS}{JS} & 0.0969793 & 0.563283 & 0.0436875 & 0.0945625 & 0.132688 & 0.199375 & 0.0241238 & 0.426079 & 0.00811287 & 0.0195179 & 0.0301587 & 0.0495591 \\
& & \expt{EN}{NO}{JS}{JS} & 0.148323 & 0.56625 & 0.0746875 & 0.156 & 0.208562 & 0.295687 & 0.0408972 & 0.429915 & 0.0146972 & 0.0356261 & 0.0527925 & 0.0873662 \\

\seprule

& \req{2} & \expt{NO}{JS}{JS}{JS} & 0.297519 & 0.565162 & 0.181875 & 0.334 & 0.422187 & 0.5395 & 0.110631 & 0.410188 & 0.0522046 & 0.113757 & 0.154439 & 0.221869 \\

\seprule

& \req{3} & \expt{EN}{JS}{JS}{JS} & \tabhvalue 0.3104654861260022 & \tabhvalue 0.5902094392776489 & 0.192125 & \tabhvalue 0.3511875 & \tabhvalue 0.4423125 & \tabhvalue 0.557875 & 0.116208 & 0.443454 & 0.0544418 & 0.119753 & 0.16262 & 0.239226 \\

\seprule

& \req{4} & \expt{NO}{TP}{JS}{JS} & 0.248357 & 0.542761 & 0.1465 & 0.276875 & 0.35 & 0.458187 & 0.0875888 & 0.399523 & 0.0394474 & 0.0868901 & 0.122163 & 0.183833 \\
& & \expt{EN}{TP}{JS}{JS} & 0.280695 & 0.568812 & 0.168563 & 0.317125 & 0.398438 & 0.513313 & 0.102004 & 0.431446 & 0.0489124 & 0.102822 & 0.141329 & 0.207525 \\
& & \expt{NO}{AL}{JS}{JS} & 0.23129 & 0.545544 & 0.133375 & 0.255375 & 0.323562 & 0.43 & 0.0784626 & 0.404235 & 0.0343327 & 0.0783069 & 0.106526 & 0.162845 \\
& & \expt{EN}{AL}{JS}{JS} & 0.310056 & 0.559869 & \tabhvalue 0.195125 & 0.348625 & 0.430312 & 0.549562 & 0.117189 & 0.41605 & 0.0556731 & 0.119988 & 0.164021 & 0.240917 \\

\seprule

& \req{5} & \exptlumem{NO}{NO}{JS}{JS} & {--} & {--} & {--} & {--} & {--} & {--} & 0.0702413 & 0.453289 & 0.0242798 & 0.0630218 & 0.0934744 & 0.154027 \\
& & \exptlumem{EN}{NO}{JS}{JS} & {--} & {--} & {--} & {--} & {--} & {--} & 0.0984506 & 0.458131 & 0.0382716 & 0.0947678 & 0.137163 & 0.213358 \\
& & \exptlumem{NO}{JS}{JS}{JS} & {--} & {--} & {--} & {--} & {--} & {--} & 0.176815 & 0.447811 & 0.0878307 & 0.190947 & 0.254145 & 0.361023 \\
& & \exptlumem{EN}{JS}{JS}{JS} & {--} & {--} & {--} & {--} & {--} & {--} & \tabhvalue 0.1842118830994603 & \tabhvalue 0.4721023895832568 & 0.092598 & \tabhvalue 0.1998353923291208 & 0.265624 & \tabhvalue 0.3763668430335097 \\
& & \exptlumem{NO}{TP}{JS}{JS} & {--} & {--} & {--} & {--} & {--} & {--} & 0.149531 & 0.43501 & 0.0718401 & 0.15485 & 0.211934 & 0.309641 \\
& & \exptlumem{EN}{TP}{JS}{JS} & {--} & {--} & {--} & {--} & {--} & {--} & 0.16722 & 0.457688 & 0.0827748 & 0.177366 & 0.240329 & 0.343857 \\
& & \exptlumem{NO}{AL}{JS}{JS} & {--} & {--} & {--} & {--} & {--} & {--} & 0.142247 & 0.440681 & 0.0663139 & 0.146855 & 0.199765 & 0.297061 \\
& & \exptlumem{EN}{AL}{JS}{JS} & {--} & {--} & {--} & {--} & {--} & {--} & 0.183997 & 0.447638 & \tabhvalue 0.09318048206937096 & 0.197119 & \tabhvalue 0.2662121929826017 & 0.373251 \\

\midrule

\java & \req{1} & \exptlu{JA} & 0.216979 & 0.498286 & 0.154214 & 0.238143 & 0.279929 & 0.340929 & 0.126734 & 0.405509 & 0.0849136 & 0.136638 & 0.166035 & 0.209229 \\
& & \exptdc{JA}{JA} & 0.158503 & 0.392351 & 0.0754286 & 0.165643 & 0.224071 & 0.3275 & 0.050319 & 0.301516 & 0.0174182 & 0.0436157 & 0.0652997 & 0.109004 \\
& & \expt{NO}{NO}{JA}{JA} & 0.0859653 & 0.544512 & 0.0352857 & 0.0807143 & 0.115857 & 0.1795 & 0.00240554 & 0.300145 & 0.000421408 & 0.00112375 & 0.00189633 & 0.00323079 \\
& & \expt{EN}{NO}{JA}{JA} & 0.125787 & 0.551958 & 0.0601429 & 0.129571 & 0.175 & 0.256 & 0.00238325 & 0.291159 & 0.000351173 & 0.00091305 & 0.00154516 & 0.00344149 \\

\seprule

& \req{2} & \expt{NO}{JA}{JA}{JA} & 0.141813 & 0.549268 & 0.0645 & 0.1475 & 0.2005 & 0.299357 & 0.0431263 & 0.409879 & 0.0146088 & 0.037435 & 0.0555556 & 0.0920073 \\

\seprule

& \req{3} & \expt{EN}{JA}{JA}{JA} & \tabhvalue 0.2907100970220285 & \tabhvalue 0.5834037660871233 & \tabhvalue 0.1732142857142857 & \tabhvalue 0.3272142857142857 & \tabhvalue 0.41700000000000004 & \tabhvalue 0.5400714285714285 & 0.114597 & 0.448746 & 0.0532378 & 0.114873 & 0.159038 & 0.239079 \\

\seprule

& \req{4} & \expt{NO}{TP}{JA}{JA} & 0.221749 & 0.528077 & 0.121286 & 0.241786 & 0.317 & 0.436929 & 0.0789662 & 0.400389 & 0.0349768 & 0.0768366 & 0.107108 & 0.163085 \\
& & \expt{EN}{TP}{JA}{JA} & 0.263584 & 0.560923 & 0.153357 & 0.291357 & 0.377929 & 0.494429 & 0.100683 & 0.434985 & 0.0464118 & 0.0992846 & 0.137341 & 0.206715 \\
& & \expt{NO}{AL}{JA}{JA} & 0.203346 & 0.530522 & 0.1085 & 0.220571 & 0.294 & 0.402643 & 0.0726539 & 0.400261 & 0.0300604 & 0.0689001 & 0.0962916 & 0.148616 \\
& & \expt{EN}{AL}{JA}{JA} & 0.284463 & 0.550411 & 0.168786 & 0.321857 & 0.402 & 0.517143 & 0.111371 & 0.419101 & 0.0507796 & 0.113116 & 0.154516 & 0.230791 \\

\seprule

& \req{5} & \exptlumem{NO}{NO}{JA}{JA} & {--} & {--} & {--} & {--} & {--} & {--} & 0.019902 & 0.369886 & 0.00393314 & 0.0114482 & 0.0186122 & 0.0345554 \\
& & \exptlumem{EN}{NO}{JA}{JA} & {--} & {--} & {--} & {--} & {--} & {--} & 0.0212444 & 0.357608 & 0.00421408 & 0.0116589 & 0.0213513 & 0.0383481 \\
& & \exptlumem{NO}{JA}{JA}{JA} & {--} & {--} & {--} & {--} & {--} & {--} & 0.109358 & 0.451518 & 0.0447394 & 0.106265 & 0.153041 & 0.236199 \\
& & \exptlumem{EN}{JA}{JA}{JA} & {--} & {--} & {--} & {--} & {--} & {--} & \tabhvalue 0.1834446770185128 & \tabhvalue 0.482121381452871 & \tabhvalue 0.0894086248068549 & \tabhvalue 0.1973591796600646 & \tabhvalue 0.266189071498806 & \tabhvalue 0.38256777637308614 \\
& & \exptlumem{NO}{TP}{JA}{JA} & {--} & {--} & {--} & {--} & {--} & {--} & 0.13957 & 0.442504 & 0.0630004 & 0.1438 & 0.201024 & 0.298834 \\
& & \exptlumem{EN}{TP}{JA}{JA} & {--} & {--} & {--} & {--} & {--} & {--} & 0.167599 & 0.471685 & 0.0794829 & 0.176078 & 0.241186 & 0.35262 \\
& & \exptlumem{NO}{AL}{JA}{JA} & {--} & {--} & {--} & {--} & {--} & {--} & 0.133059 & 0.443186 & 0.0578031 & 0.133024 & 0.187456 & 0.286487 \\
& & \exptlumem{EN}{AL}{JA}{JA} & {--} & {--} & {--} & {--} & {--} & {--} & 0.178522 & 0.458105 & 0.0867154 & 0.189212 & 0.256497 & 0.371226 \\

\midrule

\python & \req{1} & \exptlu{PY} & 0.212834 & 0.517133 & 0.152938 & 0.231687 & 0.272188 & 0.327437 & 0.121674 & 0.424811 & 0.0841454 & 0.130739 & 0.156024 & 0.194715 \\
& & \exptdc{PY}{PY} & 0.247443 & 0.444398 & 0.130125 & 0.2755 & 0.365938 & 0.505375 & 0.071806 & 0.327167 & 0.0265964 & 0.0651152 & 0.0967557 & 0.160266 \\
& & \expt{NO}{NO}{PY}{PY} & 0.109151 & 0.561703 & 0.049125 & 0.109437 & 0.150813 & 0.221312 & 0.00183205 & 0.309964 & 0.000286599 & 0.00080243 & 0.00126103 & 0.00257924 \\
& & \expt{EN}{NO}{PY}{PY} & 0.158289 & 0.580485 & 0.0790625 & 0.166438 & 0.225438 & 0.32175 & 0.00225 & 0.316154 & 0.000229279 & 0.00108907 & 0.00166217 & 0.00343918 \\

\seprule

& \req{2} & \expt{NO}{PY}{PY}{PY} & 0.235959 & 0.57233 & 0.132 & 0.260625 & 0.339313 & 0.453562 & 0.0586199 & 0.40495 & 0.0230412 & 0.0549664 & 0.0766893 & 0.124033 \\

\seprule

& \req{3} & \expt{EN}{PY}{PY}{PY} & 0.268286 & \tabhvalue 0.5941472730636597 & 0.15425 & 0.301687 & 0.386062 & 0.506563 & 0.0911606 & 0.442514 & 0.039436 & 0.0901066 & 0.12404 & 0.190989 \\

\seprule

& \req{4} & \expt{NO}{TP}{PY}{PY} & 0.247746 & 0.547862 & 0.140313 & 0.277187 & 0.354875 & 0.468625 & 0.0571079 & 0.394496 & 0.02115 & 0.0525023 & 0.0764607 & 0.121856 \\
& & \expt{EN}{TP}{PY}{PY} & 0.298995 & 0.577323 & 0.180812 & 0.337875 & 0.423187 & 0.545438 & 0.107365 & 0.429882 & 0.04998 & 0.105806 & 0.149596 & 0.223362 \\
& & \expt{NO}{AL}{PY}{PY} & 0.232225 & 0.556429 & 0.128812 & 0.256312 & 0.333875 & 0.45075 & 0.0622415 & 0.406051 & 0.0244169 & 0.057546 & 0.0832239 & 0.130338 \\
& & \expt{EN}{AL}{PY}{PY} & \tabhvalue 0.30693374258033745 & 0.568091 & \tabhvalue 0.18925 & \tabhvalue 0.34725 & \tabhvalue 0.432375 & \tabhvalue 0.5479375 & 0.112288 & 0.420689 & 0.051817 & 0.114232 & 0.158193 & 0.228922 \\

\seprule

& \req{5} & \exptlumem{NO}{NO}{PY}{PY} & {--} & {--} & {--} & {--} & {--} & {--} & 0.0226943 & 0.366935 & 0.0049295 & 0.0142153 & 0.0221814 & 0.0431044 \\
& & \exptlumem{EN}{NO}{PY}{PY} & {--} & {--} & {--} & {--} & {--} & {--} & 0.0269871 & 0.368561 & 0.00659139 & 0.018457 & 0.0292331 & 0.0524476 \\
& & \exptlumem{NO}{PY}{PY}{PY} & {--} & {--} & {--} & {--} & {--} & {--} & 0.134218 & 0.442164 & 0.0593798 & 0.1367 & 0.191093 & 0.285837 \\
& & \exptlumem{EN}{PY}{PY}{PY} & {--} & {--} & {--} & {--} & {--} & {--} & 0.172374 & \tabhvalue 0.4744940559538518 & 0.0830563 & 0.18348 & 0.249857 & 0.360025 \\
& & \exptlumem{NO}{TP}{PY}{PY} & {--} & {--} & {--} & {--} & {--} & {--} & 0.131258 & 0.433332 & 0.055368 & 0.135267 & 0.186394 & 0.279132 \\
& & \exptlumem{EN}{TP}{PY}{PY} & {--} & {--} & {--} & {--} & {--} & {--} & 0.182566 & 0.464538 & 0.09056 & 0.1935 & 0.264916 & 0.380868 \\
& & \exptlumem{NO}{AL}{PY}{PY} & {--} & {--} & {--} & {--} & {--} & {--} & 0.137449 & 0.442381 & 0.0624749 & 0.136986 & 0.193501 & 0.291626 \\
& & \exptlumem{EN}{AL}{PY}{PY} & {--} & {--} & {--} & {--} & {--} & {--} & \tabhvalue 0.18703313448771441 & 0.452249 & \tabhvalue 0.09314456035767511 & \tabhvalue 0.20020630371950643 & \tabhvalue 0.2702470446173515 & \tabhvalue 0.38235802276210873 \\

\midrule

\toplangs & \req{4} & \exptlu{TP} & 0.252115 & 0.527764 & 0.187021 & 0.275708 & 0.317937 & 0.377417 & 0.112406 & 0.39564 & 0.0770918 & 0.121285 & 0.145617 & 0.181223 \\
& & \exptdc{TP}{TP} & 0.324149 & 0.48554 & 0.205208 & 0.36625 & 0.453625 & 0.57025 & 0.0493739 & 0.303776 & 0.0182233 & 0.0444496 & 0.0654097 & 0.104168 \\
& & \expt{NO}{NO}{TP}{TP} & 0.0226522 & 0.366936 & 0.00566667 & 0.0150417 & 0.0237083 & 0.0422083 & 0.00188142 & 0.271956 & 0.00035252 & 0.00101531 & 0.00147169 & 0.00275859 \\
& & \expt{EN}{NO}{TP}{TP} & 0.0228787 & 0.391944 & 0.00595833 & 0.0157083 & 0.02375 & 0.0430625 & 0.00186132 & 0.29689 & 0.000352453 & 0.00101601 & 0.00143095 & 0.00284133 \\
& & \expt{NO}{TP}{TP}{TP} & 0.386129 & 0.611085 & 0.260979 & 0.440833 & 0.527687 & 0.639729 & 0.0990179 & 0.3911 & 0.0468549 & 0.100116 & 0.137109 & 0.199565 \\
& & \expt{EN}{TP}{TP}{TP} & \tabhvalue 0.42767310037119044 & \tabhvalue 0.6348027658462525 & \tabhvalue 0.29500000000000004 & \tabhvalue 0.49045833333333333 & \tabhvalue 0.5817916666666667 & \tabhvalue 0.6970625 & 0.115146 & 0.412154 & 0.0567513 & 0.119321 & 0.160647 & 0.229899 \\
& & \expt{NO}{AL}{TP}{TP} & 0.401003 & 0.617259 & 0.274563 & 0.458146 & 0.542021 & 0.649583 & 0.107189 & 0.396175 & 0.0525268 & 0.10917 & 0.147669 & 0.214462 \\
& & \expt{EN}{AL}{TP}{TP} & 0.4179 & 0.628924 & 0.286021 & 0.479208 & 0.571021 & 0.684604 & 0.110385 & 0.393131 & 0.0545185 & 0.11289 & 0.152823 & 0.222026 \\

\seprule

& \req{5} & \exptlumem{NO}{NO}{TP}{TP} & {--} & {--} & {--} & {--} & {--} & {--} & 0.0342281 & 0.335255 & 0.00922022 & 0.0251794 & 0.0395904 & 0.0709924 \\
& & \exptlumem{EN}{NO}{TP}{TP} & {--} & {--} & {--} & {--} & {--} & {--} & 0.0359983 & 0.355646 & 0.00932842 & 0.026151 & 0.0413228 & 0.074642 \\
& & \exptlumem{NO}{TP}{TP}{TP} & {--} & {--} & {--} & {--} & {--} & {--} & 0.182444 & 0.42952 & 0.0931703 & 0.197057 & 0.262903 & 0.372747 \\
& & \exptlumem{EN}{TP}{TP}{TP} & {--} & {--} & {--} & {--} & {--} & {--} & \tabhvalue 0.2050088712223635 & \tabhvalue 0.44779683983602425 & \tabhvalue 0.10709190530439815 & \tabhvalue 0.22334220692905193 & \tabhvalue 0.29498632015163057 & \tabhvalue 0.412828451528965 \\
& & \exptlumem{NO}{AL}{TP}{TP} & {--} & {--} & {--} & {--} & {--} & {--} & 0.190486 & 0.434495 & 0.0980442 & 0.204446 & 0.274269 & 0.384987 \\
& & \exptlumem{EN}{AL}{TP}{TP} & {--} & {--} & {--} & {--} & {--} & {--} & 0.198855 & 0.433499 & 0.102882 & 0.216002 & 0.288569 & 0.406626 \\

\bottomrule

\end{tabular}

	}
\end{table*}

\Cref{tab:results} summarizes the results of all the experiments we executed, showing the median values over \num{10} folds for all the performance metrics, \ie, \ac{mrr}, \aroma, \topk{1}, \topk{3}, \topk{5}, \topk{10}, and for both the strategies, \ie, \evalonek, \evalfull.
In the following, we will refer to \cref{tab:results}, with the addition of some plots to help the analysis.
It is worth noting that we removed \random and zero-shot results from the table and plots since their performance are close to \num{0} for all the experiments.
\begin{complete-version}
A detailed description of the result of all the experiments, together with their statistical evaluation, can be found in our online replication package~\cite{replicationpackage}.
\end{complete-version}

\subsection{Pre-Trained Single-Language Models (\req{1}--\req{3})}
\label{subsec:results:single_language}

We now describe the results of the experiments involving models that were pre-trained on English natural language or a single programming language.

\begin{figure*}[tb]
	\centering
	\includegraphics[width=1.0\linewidth]{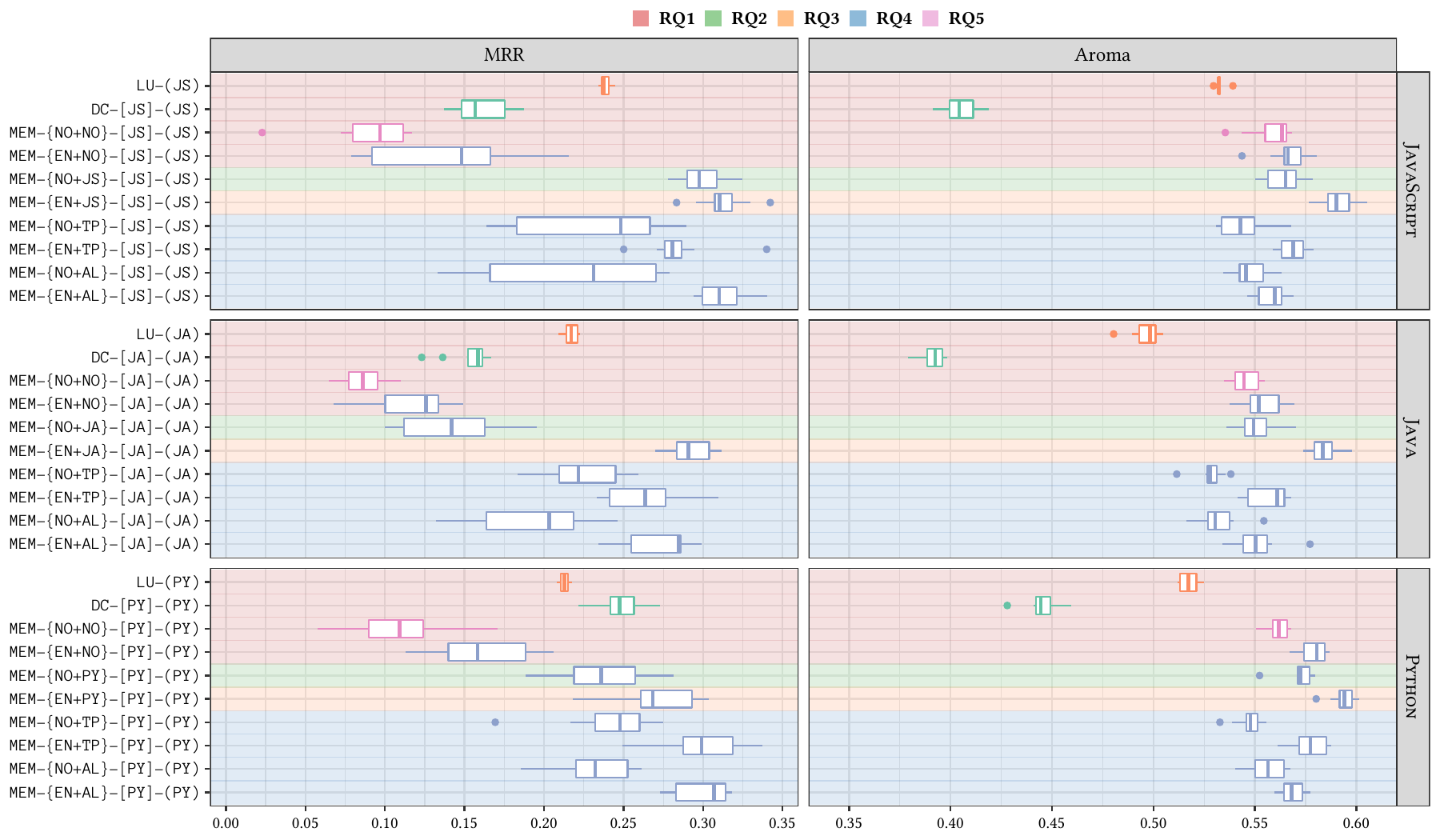}
	\caption{\ac{mrr} and \aroma values comparison for single-language test sets, using the \evalonek strategy.}
	\label{fig:jsjapy_mrr_boxplot}
\end{figure*}

\Cref{fig:jsjapy_mrr_boxplot} shows the box plots of the \num{10}-fold cross-validation \ac{mrr} and \aroma values for all the performed experiments where we use a single language as test set, considering the \evalonek strategy.
We sorted the experiments in the plots so that they could be observed in increasing order as we discuss a new research question.
We also highlight to which research questions each of the experiments refer.

Across all the experiments, the \random and zero-shot baselines, which are not presented in \cref{tab:results} and \cref{fig:jsjapy_mrr_boxplot}, reach the lowest \ac{mrr} values, close to \num{0}.
It indicates that the pre-training tasks alone are insufficient for doing code search.
Considering our model architecture, this is to be expected.
The \ac{mem} only learns \emph{unimodal} embeddings during pre-training.
However, for code search, it requires a \emph{multimodal} understanding of the data, \ie, how natural language sequences relate to source code sequences.
This relationship is only learned during the fine-tuning phase.
We address the research questions related to single-language models individually in the following.

\paragraph{Pre-trained query-only encoder (\req{1})}
\Cref{fig:jsjapy_mrr_boxplot} shows the results regarding the \evalonek strategy of the
\begin{inparaenum}[(1)]
	\item \lucene{} \exptlu{LANG},
	\item \deepcs{} \exptdc{LANG}{LANG},
	\item non-pre-trained \ac{mem} \expt{NO}{NO}{LANG}{LANG},
	\item the model with the pre-trained query encoder \expt{EN}{NO}{LANG}{LANG}, but not pre-trained on the code encoder.
\end{inparaenum}

It becomes clear that the \lucene baseline performs better than the \acp{mem} across all programming languages, reaching a median \ac{mrr} score of \num{0.2374}, \num{0.2170}, and \num{0.2128}, for \javascript{}, \java{}, and \python{}, respectively.
Compared to the non-pre-trained baselines, however, the pre-trained models show a slight improvement.
Instead, \deepcs results to be better than all the \acp{mem} for all the languages, surpassing \lucene{} only in the case of \python (\num{0.2474}).
Nevertheless, considering the \aroma score, we can notice that the \acp{mem} perform better than the others, reaching a median value of \num{0.5663}, \num{0.5520}, and \num{0.5805}, for \javascript{}, \java{}, and \python{}, respectively.
With this regard, \deepcs is considerably behind the other approaches with the median \aroma scores of \num{0.4044} (\javascript), \num{0.3924} (\java), and \num{0.4444} (\python).

In the case of the \evalfull strategy (see \cref{tab:results}), \lucene surpasses all the other approaches considering the \ac{mrr} metric: \num{0.1328}, \num{0.1267}, and \num{0.1217}, for \javascript, \java, and \python, respectively.
In particular, the query-only pre-trained model gets very slow performance in the case of \java (\num{0.0024}) and \python (\num{0.0023}) as median values for \ac{mrr}.
Instead, \aroma scores of \expt{EN}{NO}{LANG}{LANG} models and \lucene are relatively similar in the case of \javascript, \ie, \num{0.4360} and \num{0.4299}, respectively; in the other cases, \lucene gets the best results for \aroma than all the others, \ie, \num{0.4055} for \java and \num{0.4248} for \javascript.

\begin{custombox}{\req{1} -- In summary}
	The pre-trained query-only \expt{EN}{NO}{LANG}{LANG}, do not overcome the baselines of \lucene and \deepcs in terms of \ac{mrr} score, in the case of \javascript, \java, and \python, for both the \evalonek and \evalfull strategies.
	However, they get similar \aroma scores to \lucene for the \evalonek strategy.
\end{custombox}

\paragraph{Pre-trained code-only encoder (\req{2})}
We now refer to the model with the pre-trained code encoder \expt{NO}{LANG}{LANG}{LANG} in \cref{fig:jsjapy_mrr_boxplot}, \ie, we do not pre-train the query but only the code encoder.
In terms of \ac{mrr}, the code-only pre-trained models outperform the query-only pre-trained ones on all datasets when considering the \evalonek strategy.
While the pre-trained model falls behind the \lucene baseline on \java data, the pre-trained models achieve a higher median \ac{mrr} in the case of \javascript and \python.
The lower performance for \java can be explained by the smaller size of the fine-tuning dataset compared to the \javascript and \python sets (see \cref{tab:evaluation_split}).
\deepcs is still the best approach in the case of \python.
As for the \aroma score, there is no considerable difference between the query-only pre-trained models.

In the case of the \evalfull strategy, the code-only pre-trained model improves the \ac{mrr} scores for all three languages.
The code-only pre-trained model is not better in terms of \aroma than the query-only one, exclusively in the case of \javascript.
Instead, the model is statistically slightly better than \exptlu{JA}.
As for \python{}, the code-only pre-trained model is better than the query-only version.

\begin{custombox}{\req{2} -- In summary}
	The pre-trained code-only \expt{NO}{LANG}{LANG}{LANG} performs better than the pre-trained query-only \expt{EN}{NO}{LANG}{LANG}, for both the  \ac{mrr} and \aroma scores, considering both the \evalonek and \evalfull strategies.
	However, in the case of \evalonek strategy and \ac{mrr} score, \lucene remains the best model for \java and \javascript, whereas \expt{NO}{PY}{PY}{PY} performs similarly to \deepcs.
	In the case of the \evalfull strategy, the pre-trained code-only \acp{mem} still falls behind \lucene for every language.
\end{custombox}

\paragraph{Pre-trained query and code encoder (\req{3})}
We introduce the model with both the pre-trained query and code encoders \expt{EN}{LANG}{LANG}{LANG}.
As \cref{fig:jsjapy_mrr_boxplot} shows, when combining the pre-trained query encoder with code encoder \expt{EN}{LANG}{LANG}{LANG}, the \ac{mem} outperforms both the \lucene and the other baselines, including \deepcs that was the best model so far in case of \python.
This confirms our hypothesis from \req{3} that combining pre-trained models for each modality leads to even more significant improvements on code search.
Also in the case of \aroma, there is a considerable improvement for all the languages: \num{0.5902} (\javascript), \num{0.5834} (\java), and \num{0.5941} (\python).

With the \evalfull strategy, the joint contribution of both query and code pre-trained encoders improves against the versions pre-trained on one modality only, for both \ac{mrr} and \aroma scores, but not yet surpassing the performance of \lucene.

\begin{custombox}{\req{3} -- In summary}
	Pre-trained query and code \expt{EN}{LANG}{LANG}{LANG} results to be the best model in terms of \ac{mrr} and \aroma scores, in the case of the \evalonek strategy.
	However, in the case of the \evalfull strategy, \acp{mem} do not overcome the \lucene's performance.
\end{custombox}

\subsection{Pre-Trained Multi-Language Models (\req{4})}
\label{subsec:results:multi_language}

Now we present the results of the \acp{mem} that we pre-trained on several programming languages.
First, we focus on the pre-trained models on the combination of \javascript, \java, and \python data (\toplangs dataset).
Then, we show the results of the pre-trained models on the combination of \javascript, \java, \python, \php, \go, and \ruby data, \ie, the \all dataset.
While the single-language models from \cref{subsec:results:single_language} were only evaluated on single-language corpora, the experiments on multi-language models were additionally tested on the multi-language corpus \toplangs consisting of \stackoverflow question and answers for \javascript, \java, and \python.

\paragraph{Pre-trained on \toplangs dataset}
We refer to the models in \cref{fig:jsjapy_mrr_boxplot} that were pre-trained on the \toplangs dataset and evaluated on single-language corpora, using the \evalonek strategy.
We now include the models with the pre-trained encoders, \ie, \expt{NO}{TP}{LANG}{LANG}, and \expt{EN}{TP}{LANG}{LANG}.
We observe that all models in which only the code encoder was pre-trained, \eg, \expt{NO}{TP}{JS}{JS}, perform similarly or better \ac{mrr} score than their \lucene baselines.
The combined pre-trained models, \eg, \expt{EN}{TP}{JS}{JS}, improve on the \lucene baselines in all cases.
The \aroma score, instead, is never higher than the versions of \ac{mem} pre-trained on both code and query modalities but with a single language, \eg, \expt{EN}{JS}{JS}{JS}.

As for the \evalfull strategy, none of the \acp{mem} is able to surpass the performance of \lucene.
Considering the \aroma score, there is not much difference between \lucene, and other \acp{mem}.

\begin{figure*}[tb]
	\centering
	\includegraphics[width=1.0\linewidth]{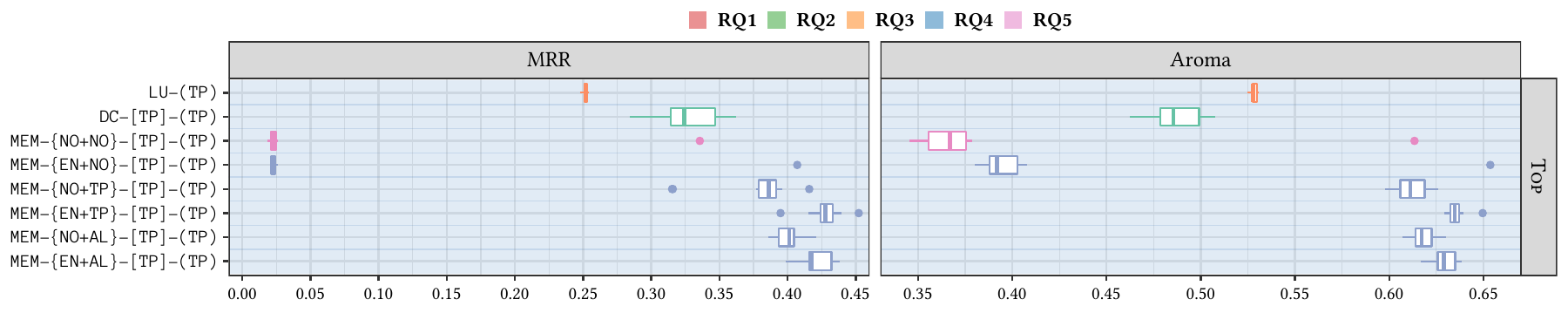}
	\caption{\ac{mrr} and \aroma values comparison for the \toplangs test set, using the \evalonek strategy.}
	\label{fig:top_mrr_boxplot}
\end{figure*}

\Cref{fig:top_mrr_boxplot} shows the box plots of the \ac{mrr} values metric for the experiments we evaluate on the \toplangs test set, considering the \evalonek strategy.
When evaluating this test set, it stands out that the pre-trained \acp{mem} outperform the baselines, when pre-training at least the code modality.
It is interesting to notice that these models considerably surpass both \lucene and \deepcs methods, performing a median value of \num{0.4277} for \ac{mrr} (\num{0.2521} \lucene, \num{0.3241} \deepcs) and \num{0.6348} for \aroma (\num{0.5278} \lucene, \num{0.4855} \deepcs).

Instead, in the case of the \evalfull strategy, the \expt{EN}{TP}{TP}{TP} model is slightly able to overcome the performance of \lucene, reaching a median value of \num{0.1151} \ac{mrr}, against \num{0.1124} of \lucene.
The \aroma media value as well is slightly better than the one performed by \lucene, \num{0.4122} instead of \num{0.3956}.

\paragraph{Pre-trained on \all dataset}
Finally, we refer to the models in \cref{fig:jsjapy_mrr_boxplot} and \cref{fig:top_mrr_boxplot} that were pre-trained on the \all dataset and evaluated on single-language search corpora.
Again, the combination of pre-training the query encoder \emph{and} the code encoder yields the best results.
These combined pre-trained models outperform both the non-pre-trained baselines and the \lucene baselines.
However, it seems that if the search is conducted on a single-language corpus, the \all dataset is not an ideal candidate for pre-training the \ac{mem}.
In this case, the better option is to pre-train on a single-language corpus of the same programming language, \eg, \expt{EN}{JS}{JS}{JS} (see also \cref{subsec:results:single_language}).
It is also confirmed by the best results for the \aroma score performed by the single language pre-trained \ac{mem}.

When looking at the results of the multi-language search for the models pre-trained on \all (\cref{fig:top_mrr_boxplot}), we observe similar results to the models pre-trained on \toplangs, both in terms of \ac{mrr} and \aroma.

As for the \evalfull strategy, we confirm the results mentioned above, suggesting overall to use an \ac{mem} pre-trained on the target language, rather than on \all.

\begin{custombox}{\req{4} -- In summary}
	Pre-trained multi-language models using \toplangs can help to improve the \ac{mrr} and \aroma performance for multi-language search, for both \evalonek and \evalfull strategies.
	However, in the case of single-language search, the use of the same target language for pre-training, \eg, \expt{EN}{JS}{JS}{JS}, results to be the best choice.
	Nevertheless, in the case of the \evalfull strategy, \lucene results to still be the best choice.
\end{custombox}

\subsection{Top-k Accuracy Values Trend}

\begin{figure}[tb]
	\centering
	\includegraphics[width=0.95\linewidth]{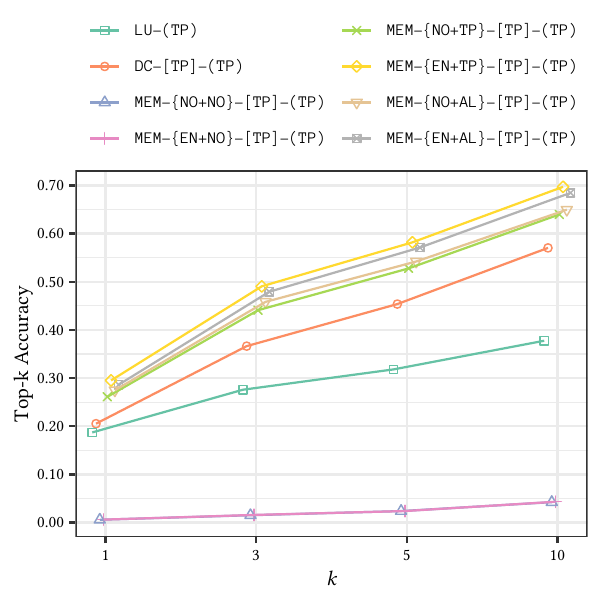}
	\caption{The trend of \topk{k} accuracy values, for the best models, using the \evalonek strategy.}
	\label{fig:topk_trend_plot}
\end{figure}

\Cref{fig:topk_trend_plot} shows the \topk{k} accuracy values for varying values of $k$.
For clarity, we only include the results of the models that were fine-tuned and evaluated on the \toplangs dataset and compare them to the \lucene baseline, considering the \evalonek strategy.
\begin{complete-version}
Refer to \cref{tab:results} and replication package~\cite{replicationpackage} for the full results.
\end{complete-version}

We observe that, with increasing values of $k$, the discrepancy in \topk{k} accuracy between the \ac{mem} and the \lucene model grows.
For $k = 10$, the best \ac{mem} puts the correct code snippet in the top results \SI{70}{\percent} of the time, while the \lucene model only does so in \SI{37}{\percent} of the cases.
We believe that expecting the user to look at \num{10} search results is acceptable, especially if the correct code snippet appears within those results with high probability.

\subsection{Combined Models (\req{5})}

\begin{figure*}[tb]
	\centering
	\includegraphics[width=1.0\linewidth]{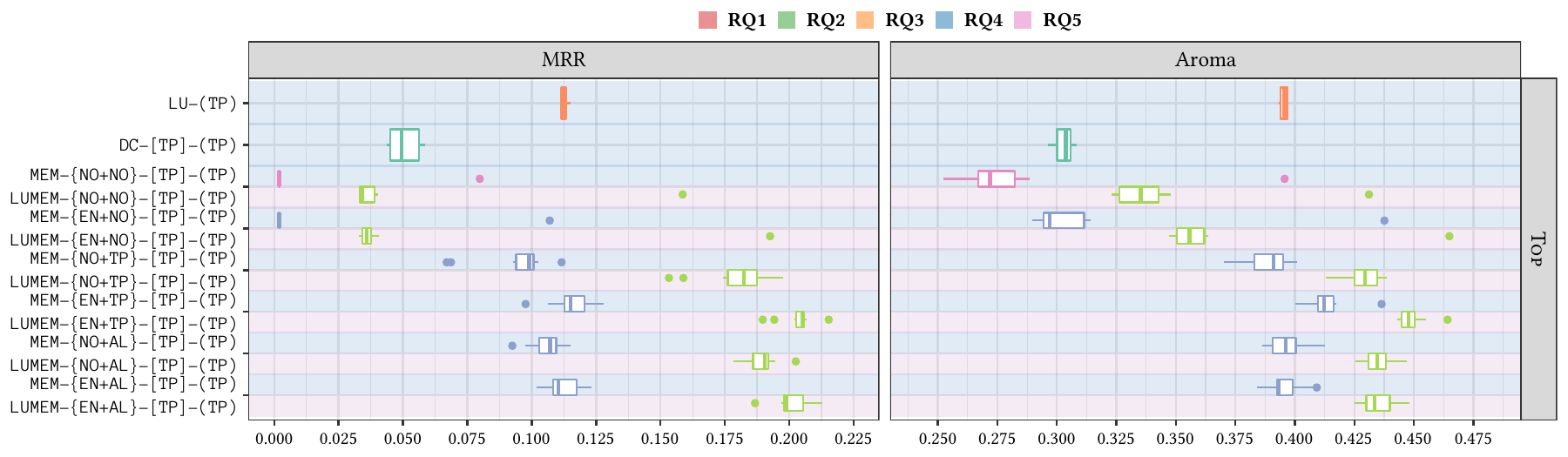}
	\caption{\ac{mrr} and \aroma values comparison for the \toplangs test set, using the \evalfull strategy, considering the combined models of \lucene and \ac{mem}.}
	\label{fig:lumem_plot}
\end{figure*}

\Cref{fig:lumem_plot} focuses on the multi-language search case (\toplangs) using the \evalfull strategy, which is the most challenging problem we addressed in our experiments, considering the size of the search space, $\num{\approx 24347}$ for each of the folds.
\begin{complete-version}
The complete results can be found in \cref{tab:results} and replication package~\cite{replicationpackage}.
\end{complete-version}
As can be seen from the figure, the combination of \lucene and \ac{mem}, \eg, \exptlumem{EN}{TP}{TP}{TP}, consinstently boosted the performance with regard to the related \acp{mem}, \eg, \expt{EN}{TP}{TP}{TP}.
More interestingly, the combined models are able to considerably overcome the \ac{mrr} performance of \lucene in the case of \toplangs languages, \ie, \num{0.1124}, reaching a median value of \num{0.2050} when pre-training on English and \toplangs, \ie, \exptlumem{EN}{TP}{TP}{TP}.
As for the \aroma score, the combined \exptlumem{EN}{TP}{TP}{TP} model now reaches the best median value of \num{0.4478}.

The same phenomena can also be verified in the case of \javascript, \java, and \python.
For the single-language search, the use of the combination of \lucene and \ac{mem}, both \ac{mrr} and \aroma values are the highest met so far.

\begin{custombox}{\req{5} -- In summary}
	The combination of \lucene and \ac{mem} has the best performance across all the programming languages, both in terms of \ac{mrr} and \aroma scores.
\end{custombox}

\subsection{Discussion}

\begin{revised}
	\Cref{tab:summary:single} and \cref{tab:summary:tp} give an overview of the results, in the case of predicting a single language, \ie, \javascript, \java, or \python, and the \toplangs languages together, respectively.
	At the top of the tables, we report the \ac{mrr} and \topk{10} absolute median values over all the experiment, for the baseline and state-of-the-art approaches, \ie, \lucene and \deepcs.
	We report the same absolute values in the case of our approach, where we did not apply any pre-train neither for the query nor code encoders, \ie, \expt{NO}{NO}{LANG}{LANG}.
	Then, we report the results for both MEM and LUMEM as median increments on top of the non-pretrained model to quantify the effectiveness of pre-training.
\end{revised}

We notice that the pre-trained models always reach higher \ac{mrr} values than the non-pre-trained ones for both the \evalonek and \evalfull strategies.
The fact that pre-training has a positive effect on code search performance means that our pre-training tasks \ac{mcm} and \ac{nlpred} are suitable pre-training tasks for source code.
The difference between the pre-trained and non-pre-trained models is particularly large when both the query encoder \emph{and} code encoder are pre-trained.
Additionally, \begin{revised}from \cref{tab:results} we can notice that\end{revised} the difference becomes more significant as the fine-tuning datasets become smaller.
We conjecture that a large fine-tuning dataset means that the model can perform well even if it was not pre-trained.
On the other hand, if the fine-tuning dataset is small, the model runs out of data before it converges to an optimum.
In this case, the pre-training can extend the training signal and allow the model to learn for longer, thus reaching a better optimum overall.

\begin{table}[tb]
	\begin{revised}
		\caption{Summary of findings for the single-language test sets}
		\label{tab:summary:single}
		\centering
		\resizebox{\linewidth}{!}{
			\sisetup{table-format=+1.4, round-mode=places, round-precision=4}
\rowcolors{2}{gray!10}{}
\begin{tabular}{
    lll SS SS
}

\hiderowcolors
\toprule

\multirow{2}[2]{*}{\textbf{Approach}} & \multirow{2}[2]{*}{\textbf{\makecell{Query\\Pre-Train}}} & \multirow{2}[2]{*}{\textbf{\makecell{Code\\Pre-Train}}} & \multicolumn{2}{c}{\textbf{1K}} & \multicolumn{2}{c}{\textbf{Full}} \\
\cmidrule(lr){4-5} \cmidrule(lr){6-7}
& & & {\textbf{\ac{mrr}}} & {\textbf{\topk{10}}} & {\textbf{\ac{mrr}}} & {\textbf{\topk{10}}} \\

\midrule
\showrowcolors

\lucene & {--} & {--} & 0.217433 & 0.340929 & 0.126734 & 0.208117 \\
\deepcs & {--} & {--} & 0.164973 & 0.337786 & 0.051782 & 0.112446 \\

\midrule

No pre-train & {--} & {--} & 0.092428 & 0.191688 & 0.002406 & 0.003371 \\

\midrule

MEM & {--} & Single & +0.1435309130 & +0.2618750000 & +0.0605503218 & +0.1309523568 \\
& & Top & +0.1505066438 & +0.2651160714 & +0.0740738179 & +0.1544022274 \\
& & All & +0.1267555231 & +0.2403035714 & +0.0651015053 & +0.1369096271 \\

\seprule

& English & {--} & +0.0465375292 & +0.0908125000 & +0.0001036932 & +0.0005838013 \\
& & Single & +0.2031102503 & \tabhvalue +0.3538660714285714 & +0.1080792278 & +0.2245459487 \\
& & Top & +0.1882671273 & +0.3245267857 & +0.1003622057 & +0.2091087177 \\
& & All & \tabhvalue +0.2038940959598621 & +0.3451875000 & +0.1102179046 & +0.2279163804 \\

\midrule

LUMEM & {--} & {--} & {--} & {--} & +0.0228570931 & +0.0451785508 \\
& & Single & {--} & {--} & +0.1318121382 & +0.2831832561 \\
& & Top & {--} & {--} & +0.1390417401 & +0.2965526028 \\
& & All & {--} & {--} & +0.1344461645 & +0.2882550253 \\

\seprule

& English & {--} & {--} & {--} & +0.0256999732 & +0.0534899116 \\
& & Single & {--} & {--} & +0.1796897448 & \tabhvalue +0.3740336089096617 \\
& & Top & {--} & {--} & +0.1675430273 & +0.3523434939 \\
& & All & {--} & {--} & \tabhvalue +0.18025848307936795 & +0.3691227961 \\

\bottomrule

\end{tabular}

		}
	\end{revised}
\end{table}

\begin{table}[tb]
	\begin{revised}
		\caption{Summary of findings for the \toplangs test set}
		\label{tab:summary:tp}
		\centering
		\resizebox{\linewidth}{!}{
			\sisetup{table-format=+1.4, round-mode=places, round-precision=4}
\rowcolors{2}{gray!10}{}
\begin{tabular}{
    lll SS SS
}

\hiderowcolors
\toprule

\multirow{2}[2]{*}{\textbf{Approach}} & \multirow{2}[2]{*}{\textbf{\makecell{Query\\Pre-Train}}} & \multirow{2}[2]{*}{\textbf{\makecell{Code\\Pre-Train}}} & \multicolumn{2}{c}{\textbf{1K}} & \multicolumn{2}{c}{\textbf{Full}} \\
\cmidrule(lr){4-5} \cmidrule(lr){6-7}
& & & {\textbf{\ac{mrr}}} & {\textbf{\topk{10}}} & {\textbf{\ac{mrr}}} & {\textbf{\topk{10}}} \\

\midrule
\showrowcolors

\lucene & {--} & {--} & 0.252115 & 0.377417 & 0.112406 & 0.181223 \\
\deepcs & {--} & {--} & 0.324149 & 0.57025 & 0.049374 & 0.104168 \\

\midrule

No pre-train & {--} & {--} & 0.022652 & 0.042208 & 0.001881 & 0.002759 \\

\midrule

MEM & {--} & Top & +0.3634769087 & +0.5975208333 & +0.0971365220 & +0.1968061224 \\
& & All & +0.3783510317 & +0.6073750000 & +0.1053078941 & +0.2117032995 \\

\seprule

& English & {--} & +0.0002265093 & +0.0008541667 & -0.0000201016 & +0.0000827372 \\
& & Top & \tabhvalue +0.40502094949483364 & \tabhvalue +0.6548541666666667 & +0.1132647470 & +0.2271408536 \\
& & All & +0.3952477369 & +0.6423958333 & +0.1085033559 & +0.2192677569 \\

\midrule

LUMEM & {--} & {--} & {--} & {--} & +0.0323467026 & +0.0682338323 \\
& & Top & {--} & {--} & +0.1805629891 & +0.3699886415 \\
& & All & {--} & {--} & +0.1886049679 & +0.3822283111 \\

\seprule

& English & {--} & {--} & {--} & +0.0341168981 & +0.0718834267 \\
& & Top & {--} & {--} & \tabhvalue +0.20312744671774124 & \tabhvalue +0.41006985814670643 \\
& & All & {--} & {--} & +0.1969732553 & +0.4038678121 \\

\bottomrule

\end{tabular}

		}
	\end{revised}
\end{table}

Moreover, we noticed that pre-training the query encoder on natural language is smaller than the effect of pre-training the code encoder on source code, even though the query encoder was pre-trained for longer and on a larger dataset than the code encoder.
We attribute this difference to the fact that queries tend to be much shorter (around \num{9} tokens) than the code snippets (around \num{225} tokens, see \cref{tab:so_quality}).
\bert and the pre-training tasks are designed to learn the relationships between tokens in a sequence, particularly for distant token pairs.
The shorter the sequence is, the less impactful the learned contextual embeddings from the pre-training become as there is less context information that the model can utilize.

Considering the number of experiments we performed, we can use the results from both the \evalonek and \evalfull strategies to derive the following observations.
The experiments conducted using the \evalonek strategy, \ie, reducing the search pool for each of the queries to \num{1000} samples from the test set, suggest that the \acp{mem} based on both code and query pre-trained encoders, \ie, \expt{EN}{LANG}{LANG}{LANG}, allow outperforming the state of the art, \ie, \deepcs, and the information retrieval-based approach, \ie, \lucene.
The \ac{mrr} scores are sufficiently high for all the programming languages, including the multi-language one, \ie, \toplangs.
The \topk{k} values indicate that \acp{mem} can rank as first in the correct code snippets for $\SI{\approx 20}{\percent}$ of the cases, up to $\SI{\approx 60}{\percent}$ of the cases within the first \num{10} results.\
Furthermore, the \aroma scores suggest that \acp{mem} are also effective in finding alternative code snippets which are still semantically close to the given search queries.
Overall, this indicates that \acp{mem} are the most effective approaches in ranking code snippets when the search space is \num{1000}.

Instead, when considering the \evalfull strategy, \ie, the entire test dataset for each of the search target programming languages, the \acp{mem} do not overcome the performance of the information retrieval approach, \ie, \lucene.
However, by producing a model as the combination of \lucene and \ac{mem}, we could get a considerable boost in all the analyzed performance metrics.
It suggests using such a combined approach as a pipeline for enhanced search engines:
\begin{inparaenum}[(1)]
	\item the \lucene approach is first run to reduce the size of the search pool.
	In our experiments, we used \num{1000} since we already collected some evidence of the effectiveness of \acp{mem} when considering the \evalonek strategy.
	These values could be further tuned.
	\item Once a set of candidates has been selected, the search engine can run \ac{mem} on it to propose the best results.
\end{inparaenum}

\section{Related Work}
\label{sec:related_work}

The main goal of our work is to show that using transfer learning in the form of pre-trained source code models is beneficial to code search performance.
While other approaches use more sophisticated models and may even outperform ours, we hope to motivate the use of transfer learning, \eg, by using one of our pre-trained models as a starting point.
In the following, we list the current state-of-the-art approaches for deep learning tailored for code intelligence, \ie, problems involving source code, and a focus on code search.

\subsection{Deep Learning for Code Intelligence}

The literature presents several contributions of deep learning to solve code-related problems~\cite{lu_codexglue_2021}, \ie, problems in which source code is treated as input.
In theory, our pre-trained models could be fine-tuned to solve also the following tasks.

\emph{Clone detection} consists in measuring the semantic between codes~\cite{mou_convolutional_2016, svajlenko_big_2014}.
\emph{Defect detection} aims at identifying whether code contains defects~\cite{zhou_devign_2019}.
\emph{Code completion} expects to predict the following tokens based on the current context~\cite{allamanis_mining_2013, raychev_probabilistic_2016}.
\emph{Code translation} involves the conversion from a programming language to another~\cite{nguyen_divideandconquer_2015}.
\emph{Code repair} aims at fixing bugs in the code automatically~\cite{tufano_empirical_2019}.
\emph{Code generation} has the goal of creating code automatically on the base of ta provided natural language description~\cite{iyer_mapping_10}, whereas \emph{code summarization} aims at describing with text a piece of code~\cite{iyer_summarizing_2016}.

\paragraph{Code representation}
Code data can be treated by models in different ways.
In the case of token sequences as in \ac{nlp}, self-supervised representation learning can be tailored for code data, such as \textsc{CodeBERT}~\cite{feng_codebert_2020}, \textsc{Codex}~\cite{chen_evaluating_2021}, and \textsc{PLBART}~\cite{ahmad_unified_2021}.
Instead, \textsc{ContraCode}~\cite{jain_contrastive_2020} builds representations of program functionalities by learning from contrastive samples~\cite{hadsell_dimensionality_2006}.

Also, it is common first to parse the code into tree or graph structures also to catch the semantics.
\textsc{ASTNN}~\cite{zhang_novel_2019} splits each \ac{ast} into a sequence of small trees for better representations.
\textsc{MRNCS}~\cite{gu_multimodal_2021} recaps serialization schemes on tree structures and categorized them into sampling-based~\cite{alon_general_2018}, and traversal-based ones~\cite{hu_deep_2018}.
\textsc{TDLS}~\cite{allamanis_learning_2018} uses \textsc{GGNN}~\cite{li_gated_2016} to learn both syntactic and semantic information. 
Instead, \textsc{DyPro}~\cite{wang_learning_2019a} and \textsc{LiGer}~\cite{wang_learning_2019} learn program representations through dynamic executions, from the mixture of symbolic and concrete execution traces.
Zhang \etal~\cite{zhang_disentangled_2021} addressed the problem of code representation using a multi-language setting to create an embedding that separates the semantic from the context of the source code.
\begin{revised}
\textsc{Flow2Vec}~\cite{sui_flow2vec_2020} is an embedding approach that preserves interprocedural program dependence by approximating the high-order proximity.
\end{revised}

Even though we use programming languages encoders in our work, our approach uses code snippets that are hardly parsable, \ie, the construction of an \ac{ast} is not possible.
As explained in \cref{sec:approach}, our approach is token-based.

\subsection{Code Search}

The code search problem aims at measuring the semantic relation between a text and source code~\cite{lu_codexglue_2021,husain_codesearchnet_2019}.
In the following, we describe the literature work that uses machine learning to address the problem.

\paragraph{Code search using non-\transformer models}
Allamanis \etal~\cite{allamanis_bimodal_2015} learn bimodal representations of source code and natural language and apply them to code search.
Iyer \etal\cite{iyer_summarizing_2016} extend the work of Allamanis \etal by training a \ac{lstm} neural network with attention and applying it to code summarization, \ie, generating natural language descriptions from code snippets.
Ye \etal~\cite{ye_word_2016} learn token embeddings based on the Skip-gram model by Mikolov \etal~\cite{mikolov_distributed_2013}.
They evaluate their embeddings on two code search tasks: bug localization, \ie, given a bug report, find affected source code files, and \ac{api} recommendation for \stackoverflow questions.

Sachdev \etal~\cite{sachdev_retrieval_2018} derive a purely token-based approach to code search.
They use the \textsc{fastText} algorithm~\cite{bojanowski_enriching_2017} to learn embeddings for source code tokens.
Then, they use these embeddings to encode both the source code and the search query.
Wan \etal~\cite{wan_multimodal_2019} combine multiple source code representations: source code tokens, \acp{ast}, and control flow graphs.
By using attention, they hypothesize that the neural network will automatically select the most valuable features from the different representations.
Finally, Gu \etal~\cite{gu_deep_2018} propose \deepcs, a neural network-based approach that uses two independent \ac{lstm} models for the representation of code snippets and queries.
\deepcs can compute the vector representations for both the query and code, then use a similarity function to find the best match.
In our work, we chose \deepcs as the representative state-of-the-art neural model for code search (\cref{subsec:experimental_design:baselines}).

\paragraph{Code search using transformers}
Husain \etal~\cite{husain_codesearchnet_2019} build a range of neural network models and compare their performance on the code search task.
One of their models is the \bert-based \emph{self-attention} model.
They train and evaluate their models on pairs of docstring-code pairs mined from open-source repositories on \github.
Even without pre-training, the self-attention model shows good results when trained and evaluated on the same dataset.
They also evaluate their models on a manually annotated dataset.
Our work closely resembles that of Husain \etal, as we use the same multimodal embedding architecture (see \cref{subsec:background:mem}).
The main difference with their approach is that we \emph{pre-train} the encoders before applying them to code search, \ie, we use the concept of transfer learning.
Furthermore, while we also use \github data for pre-training, we fine-tune and evaluate our model on \stackoverflow data, which we believe better approximates code search than docstring-code pairs.

Feng \etal~\cite{feng_codebert_2020} build on the work of Husain \etal, but instead of using cosine similarity between the outputs of two separate encoders, they concatenate the query and code sequence, feed it to a \emph{single} encoder, and measure the similarity between query and code snippet by using a summarizing token in the output sequence.
While using a single encoder reduces model complexity, our two-encoder architecture offers more flexibility by allowing various combinations of pre-trained natural language and source code models.
Moreover, we are interested in anatomically analyzing the impact of pre-trained modalities; therefore, we need to treat the encoders separately.

Shuai \etal~\cite{shuai_improving_2020} employ a similar architecture to ours, which they call \textsc{CARLCS-CNN}.
It uses the \emph{co-attention} method to build the semantic relationship between code snippets and related queries.
Similarly, Fang \etal~\cite{fang_selfattention_2021} proposed \textsc{SAN-CS}, which is solely based on the \emph{self-attention} method to achieve the same purpose.
While transfer learning would also be possible with their architecture, it is not discussed in their work.
Their work also differs from ours in that they use docstrings as query proxies while we use \stackoverflow question titles.
Even though docstrings are commonly used as code search queries, we believe that \stackoverflow question titles are a better approximation to real-world queries.

Although \textsc{CARLCS-CNN}~\cite{fang_selfattention_2021} and \textsc{SAN-CS}~\cite{shuai_improving_2020} claimed better performance, they bring additional restrictions to the compatibility with our experiments.
Both \textsc{CARLCS-CNN}~\cite{fang_selfattention_2021} and \textsc{SAN-CS}~\cite{shuai_improving_2020} introduce the co-attention mechanism to refine the code and query representations.
Other than generating the independent vector representations for the code snippet and query, they compute a joint attention representation, aiming to catch the semantic information and the semantic relation between the two parts.
During the prediction phase, \ie, the search, the approaches
\begin{inparaenum}[(1)]
    \item compute the independent representations for all the queries and code snippets,
    \item compute the pairwise joint representations between all the possible queries and snippets,
    \item for each query, select the best matching code snippet. 
\end{inparaenum}
This mechanism is better suitable for a \emph{non}-blind code search problem, \ie, the sets of queries and code snippets are known a priori, and the approach is asked to find the correct matching pairs.
Instead, we are interested in a blind search that better simulates an arbitrary input of a user: the user writes their query, and the approach looks for the best match in the entire code base.
It is the main reason why we compare with \deepcs~\cite{gu_deep_2018}.

\section{Conclusions and Future Work}
\label{sec:conclusions}

We demonstrated that transfer learning is an effective method for improving code search performance of neural networks.
The impact of transfer learning is particularly noticeable in cases where limited training data is available.
Because many code intelligence problems are limited by the size of the training dataset and that large code corpora can easily be obtained from open source platforms such as \github, we advocate that transfer learning can lead to improvements also for other source code analysis tasks.

We showed that state-of-the-art sequence-to-sequence models such as \bert that were initially designed for \ac{nlp} tasks can successfully be applied to problems dealing with source code data.
However, due to many parameters of such models, they require extensive amounts of pre-training and fine-tuning data.
In cases where both these training sets are small, a \lucene model achieves similar or better results in code search.
However, we demonstrated that the combined use of an information retrieval approach, \ie, \lucene, followed by a pre-trained \ac{mem} on a filtered amount of search candidates, brings the best performance in terms of \ac{mrr} and \aroma score values.

Moreover, we found some evidence that \bert, while being effective at modeling long sequences with hundreds of tokens, may be limited in modeling concise ones (fewer than \num{10} tokens).
As search queries tend to be short, this might be a limiting factor of \bert when applied to code search.

Despite these findings, there are still open questions to address in the future.
Our code encoder treats source code the same as natural language, namely as a sequence of tokens.
While we have demonstrated that such a token-based model can yield good results on code search, we expect it to perform even better if the model uses the highly structured nature of source code.
It can be achieved, for example, by replacing or augmenting the token-based input to the code encoder with input features representing the structural information of source code, \eg, \acp{ast}.

Moreover, other preprocessing operations might be applied to optimize the performance of the presented models, \eg, using formatters to normalize the code before processing it.
In the case of pre-training on multiple programming languages, we intend to investigate the use of specific optimizations like the cross-lingual language model pre-training~\cite{conneau_crosslingual_2019}.

Finally, it would be insightful to inspect \bert's attention heads when processing source code similarly to natural language, for which it has been shown that the attention heads focus on specific language constructs, \eg, verbs and their objects~\cite{clark_what_2019}.
Such understanding of the model's inner workings can drive the development of better model architectures for code search and other source code analysis tasks.

\section*{Acknowledgements}

We are grateful for the anonymous reviewers' comments and feedback that helped to improve the paper significantly.
The research leading to these results has received funding from the Swiss National Science Foundation (SNSF) project \enquote{Melise - Machine Learning Assisted Software Development} (SNSF204632).

\bibliography{references, urls}

\begin{complete-version}
\begin{IEEEbiography}%
	[{\includegraphics[width=1in,height=1.25in,clip,keepaspectratio]{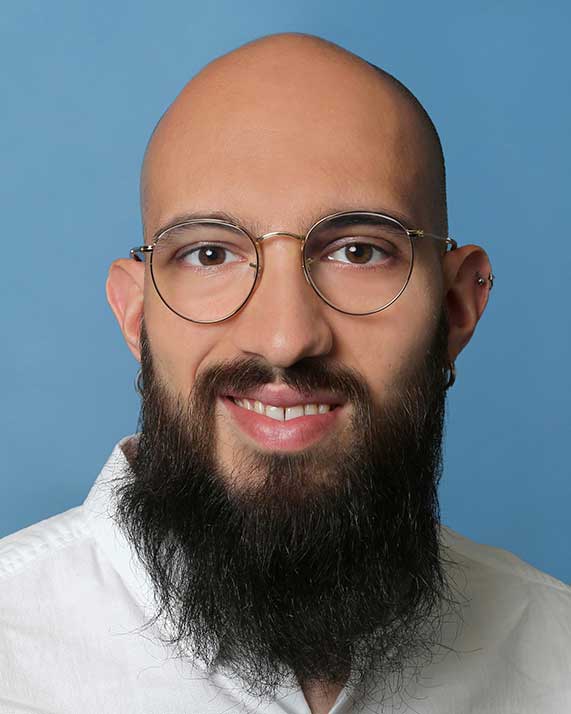}}]
	{Pasquale Salza}%
	is a Senior Research Associate
    in the Software Evolution and Architecture Lab (s.e.a.l.)
    at the University of Zurich, Switzerland.
    He received a Ph.D. degree in Computer Science from the University of Salerno, Italy.
	His research interests include software engineering, machine learning, cloud computing, and evolutionary computation.
	Contact him at \href{mailto:salza@ifi.uzh.ch}{salza@ifi.uzh.ch}.
\end{IEEEbiography}

\begin{IEEEbiography}%
	[{\includegraphics[width=1in,height=1.25in,clip,keepaspectratio]{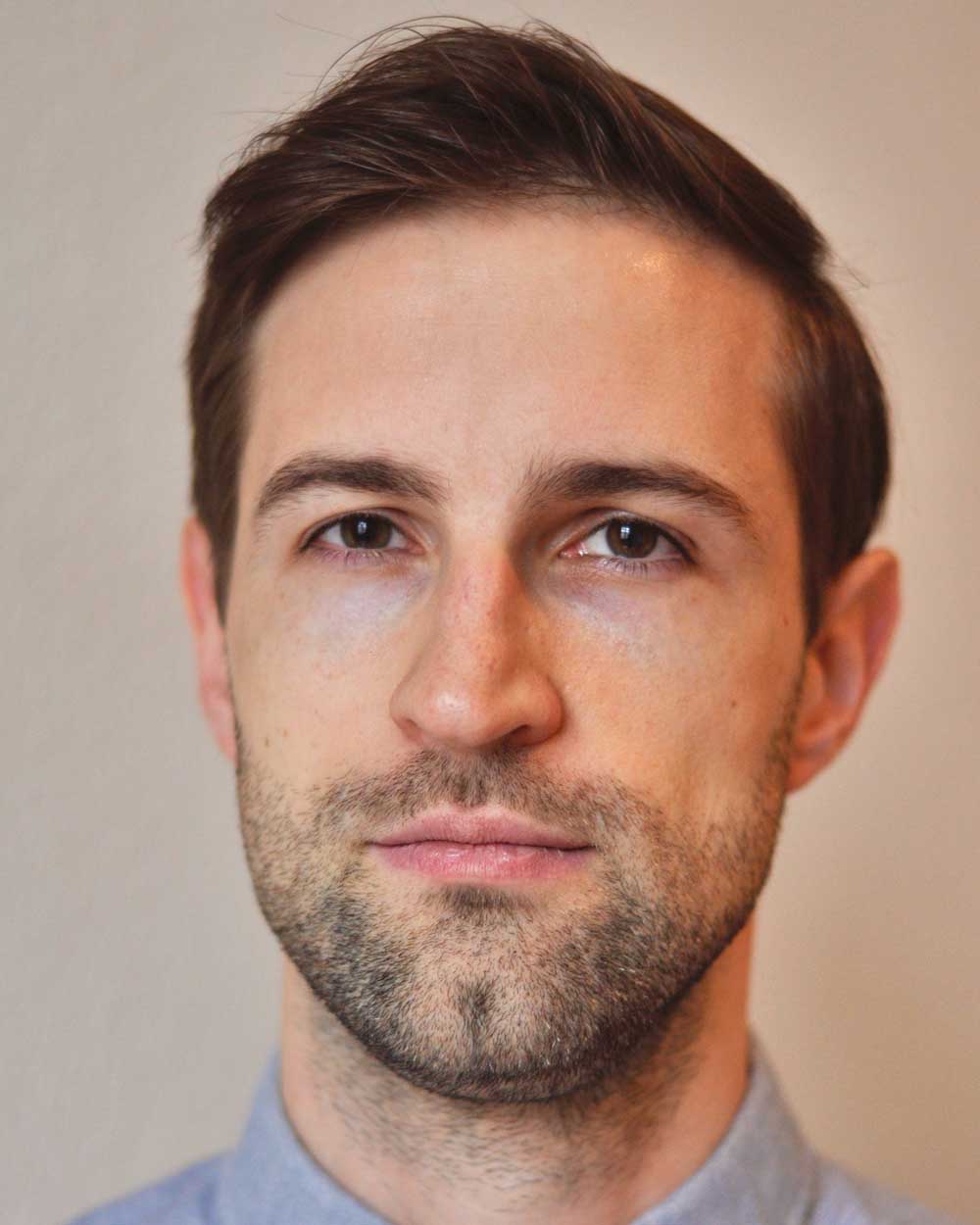}}]
    {Christoph Schwizer}%
    received the M.Sc. degree in Computer Science from the University of Zurich, Switzerland.
    His research focused on the use of deep learning models to solve software engineering problems.
    He currently works as a software engineer in the digitization of public sector services, in particular electronic identity.
    Contact him at \href{mailto:christoph@schwizer.dev}{christoph@schwizer.dev}.
\end{IEEEbiography}

\begin{IEEEbiography}%
	[{\includegraphics[width=1in,height=1.25in,clip,keepaspectratio]{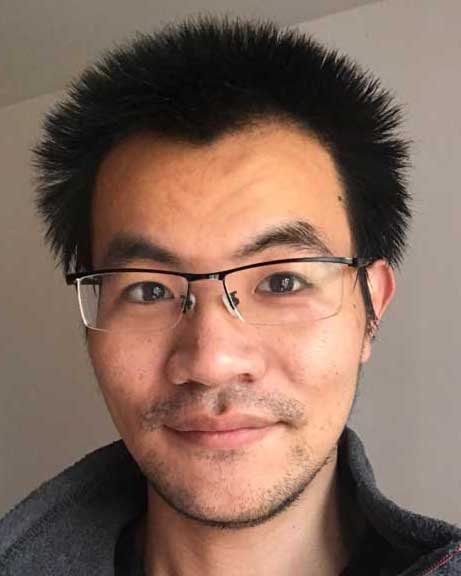}}]
    {Jian Gu}%
    received the M.Sc. degree in Machine Learning from KTH Royal Institute of Technology, Sweden.
    He is currently working toward the Ph.D. degree in Computer Science at the Department of Informatics, University of Zurich, Switzerland,
    and a member of the Software Evolution and Architecture Lab (s.e.a.l.).
    His research interests include software engineering and machine learning.
    Contact him at \href{mailto:gu@ifi.uzh.ch}{gu@ifi.uzh.ch}.
\end{IEEEbiography}

\begin{IEEEbiography}%
	[{\includegraphics[width=1in,height=1.25in,clip,keepaspectratio]{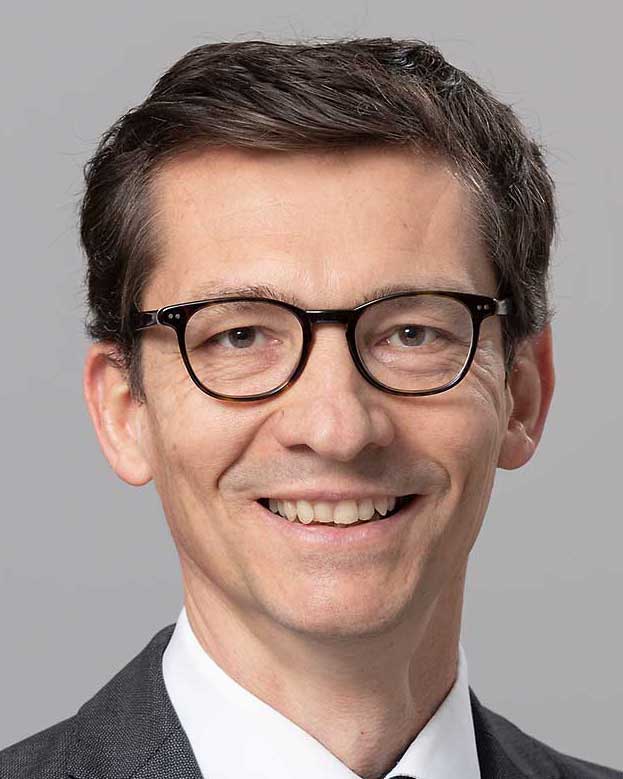}}]
    {Harald C. Gall}%
    is Dean of the Faculty of Business, Economics, and Informatics at the University of Zurich.
    He is professor of software engineering in the Department of Informatics.
    He held visiting positions at Microsoft Research in Redmond, USA, and University of Washington in Seattle, USA.
    His research interests are software evolution, software architecture, software quality, and cloud-based software engineering.
    Since 1997, he has worked on devising ways in which mining repositories can help to better understand and improve software development.
    Contact him at \href{mailto:gall@ifi.uzh.ch}{gall@ifi.uzh.ch}.
\end{IEEEbiography}

\vfill

\end{complete-version}

\end{document}